\documentclass[aps,showpacs,twocolumn,nofootinbib]{revtex4}

\usepackage{graphicx}
\usepackage{epstopdf}
\usepackage{dcolumn}
\usepackage{amsmath}

\begin{document}
\title{Symmetries of the Interacting Boson Model}

\author{P.~Van~Isacker}

\affiliation{Grand Acc\'el\'erateur National d'Ions Lourds, CEA/DRF--CNRS/IN2P3\\
Bd Henri Becquerel, BP 55027, F-14076 Caen, France}

\date{\today}

\begin{abstract}
In this contribution the symmetry properties of the interacting boson model of Arima and Iachello are reviewed.
While the concept of a dynamical symmetry is by now a familiar one,
this is not necessarily so for the extended notions
of partial dynamical symmetry and quasi dynamical symmetry,
which can be beautifully illustrated
in the context of the interacting boson model.
The main conclusion of the analysis is that
dynamical symmetries are scarce
while their partial and quasi extensions are ubiquitous.
\end{abstract}

\maketitle

\section{Introduction}
\label{s_intro}
In 1975 Arima and Iachello proposed a new approach to nuclear collective motion,
the interacting boson model or IBM~\cite{Arima75}.
It quickly became a popular model for the interpretation of nuclear data
and acquired the center stage of discussions within the nuclear-structure community
in the remainder of the decade and much of the 1980s.

What made and still makes the model so appealing?
One of its strengths is that it offers a unified view of several descriptions
which until the 1970s existed more or less separately.
Nuclei can be viewed as incompressible, charged liquid drops,
which vibrate and, if deformed, also rotate~\cite{Rainwater50,Bohr52}.
From this picture are derived a variety of models
such as the anharmonic spherical vibrator~\cite{Brink65},
the deformed rotor-vibrator~\cite{Bohr53},
or the $\gamma$-unstable rotor~\cite{Wilets56}.
The IBM includes all three descriptions
as special cases of its Hamiltonian.
Not only do such cases turn out to be analytically solvable
through the use of dynamical symmetries
but, in addition, one may easily interpolate in the IBM
between the different geometric solutions.

A second advantage of the IBM
concerns its connection with the nuclear shell model.
Although originally proposed as purely phenomenological description of nuclei,
it was realized early on that a microscopic justification can be obtained
by considering the bosons as pairs of valence nucleons 
with angular momentum $L=0$ or $L=2$~\cite{Arima77,Otsuka78a}.
The model therefore has acquired a status
that is intermediate between phenomenological collective
and more microscopic single-particle models.
This shell-model interpretation of the IBM
also showed the direction of various extensions,
extendibility being another one of its strong points.
Given the interpretation of the bosons as pairs of nucleons,
it was soon suggested that bosons should come in two varieties,
neutron and proton, giving rise to the \mbox{IBM-2}~\cite{Otsuka78b}.
Furthermore, Elliott and co-workers~\cite{Elliott80,Elliott81}
pointed out that the existence of corresponding symmetries
in the interacting boson model and in the nuclear shell model
can be put to good use in the microscopic justification of the former.

But perhaps the most characteristic aspect of the IBM
is its symmetry-based formulation.
Symmetries, of central importance to physics,
also play a pivotal role in the IBM,
which makes in particular extensive use of the notion of dynamical symmetry.
While the latter might still be unfamiliar as a term,
it appears in diverse areas of physics,
also in nuclear physics where it received wide-spread attention.
Notable examples are Wigner's supermultiplet model~\cite{Wigner37},
Racah's pairing model~\cite{Racah43},
and Elliott's rotation model~\cite{Elliott58},
and their many extensions that can be formulated
in terms of dynamical symmetries~\cite{Iachello06,Frank09}.

It is not the aim of this contribution
to give a comprehensive review of the properties of the IBM
but rather to focus on the use of notions of symmetry in the model.
It specifically deals with two further generalizations
of the concept of dynamical symmetry,
which have been developed over the last two decades,
namely partial dynamical symmetry and quasi dynamical symmetry.
Again, the IBM proved to be instrumental
in the development of these novel symmetry notions
and the basic ideas behind these extensions
can be illustrated beautifully with a simplified IBM Hamiltonian~\cite{Isacker14}.
Before turning to these extended notions of symmetry,
a brief review of the IBM is given.

\section{The interacting boson model}
\label{s_ibm}
\subsection{Basic properties}
\label{ss_basic}
In the original version of the IBM as applied to even-even nuclei,
collective properties of the nucleus
are described in terms of a set of interacting $s$ and $d$ bosons
carrying the angular momenta $\ell=0$ and $\ell=2$, respectively.
In the simplest version of the model, referred to as \mbox{IBM-1},
it is assumed that there is only one kind of boson
(i.e., no distinction is made between neutron and proton bosons)
and that they carry no further intrinsic labels such as spin or isospin.
The associated creation and annihilation operators
satisfy the standard boson commutation relations
\begin{align}
[b_{\ell m},b^\dag_{\ell'm'}]&=\delta_{\ell\ell'}\delta_{mm'},
\nonumber\\
[b_{\ell m},b_{\ell'm'}]&=[b^\dag_{\ell m},b^\dag_{\ell'm'}]=0.
\label{e_bosoncom}
\end{align}
The \mbox{IBM-1} assumes that low-lying collective states of an even-even nucleus
can be described in terms of boson excitations
acting upon a vacuum state $|{\rm o}\rangle$,
which is interpreted as the doubly-closed core of the nucleus under consideration.
There are six basic excitations,
$s^\dag|{\rm o}\rangle$ and $d^\dag_m|{\rm o}\rangle,m=0,\pm1,\pm2$,
and the unitary transformations among them generate the Lie algebra U(6).
A different way of expressing the same property
is through the construction of the bilinear operators $b_{\ell m}^\dag b_{\ell'm'}$,
which generate U(6)~\cite{Iachello06}.

As mentioned in the introduction,
the bosons are associated with (collective) pairs of nucleons in the valence shell.
Because of this interpretation,
a collective state of an even-even nucleus with $2N_{\rm b}$ valence nucleons
is approximated as a state with $N_{\rm b}$ bosons.
In general, the separate boson numbers $n_s$ and $n_d$ are not conserved
but the sum $n_s+n_d=N_{\rm b}$ is.
The Hamiltonian of the \mbox{IBM-1}
can therefore be written in terms of the generators of the Lie algebra U(6)
and acquires the generic form
\begin{equation}
\hat H=E_0+\hat H_1+\hat H_2+\hat H_3+\cdots,
\label{e_ham}
\end{equation}
where the index refers to the order of the interaction in the generators of U(6).
The first term $E_0$ is a constant
and represents the (negative of the) binding energy of the doubly-closed core. 
The second term is the one-body part
\begin{align}
\hat H_1&{}=
\epsilon_s[s^\dag\tilde s]^{(0)}+
\epsilon_d\sqrt{5}[d^\dag\tilde d]^{(0)}
\nonumber\\&{}=
\epsilon_s\, s^\dag\cdot\tilde s+
\epsilon_d\, d^\dag\cdot\tilde d\equiv
\epsilon_s\,\hat n_s+
\epsilon_d\,\hat n_d,
\label{e_ham1}
\end{align}
where the coupling in angular momentum is shown as an upperscript in round brackets
and the dot indicates a scalar product.
Furthermore, $\tilde b_{\ell m}\equiv(-)^{\ell-m}b_{\ell,-m}$
and the coefficients $\epsilon_s$ and $\epsilon_d$
are the single-boson energies in the $s$ and $d$ state, respectively.
The third term in the Hamiltonian~(\ref{e_ham}) is the two-body interaction,
\begin{equation}
\hat H_2=
\sum_{\ell_1\leq\ell_2,\ell'_1\leq\ell'_2,L}
\tilde v^L_{\ell_1\ell_2\ell'_1\ell'_2}
[b^\dag_{\ell_1}b^\dag_{\ell_2}]^{(L)}\cdot[\tilde b_{\ell'_2}\tilde b_{\ell'_1}]^{(L)},
\label{e_ham2}
\end{equation}
where the coefficients $\tilde v^L_{\ell_1\ell_2\ell'_1\ell'_2}$ are related to the interaction matrix elements
between normalized two-boson states, 
\begin{equation}
\tilde v^L_{\ell_1\ell_2\ell'_1\ell'_2}
=(-)^L\frac{\langle\ell_1\ell_2;L|\hat H_2|\ell'_1\ell'_2;L\rangle}{\sqrt{(1+\delta_{\ell_1\ell_2})(1+\delta_{\ell'_1\ell'_2})}}.
\end{equation}
The bosons are symmetrically coupled
and allowed two-boson states are:
$s^2$  with angular momentum $L=0$,
$sd$ with $L=2$, and $d^2$ with $L=0,2,4$.
This leads to seven independent two-body interactions:
three for $L=0$, three for $L=2$, and one for $L=4$.

\begin{table}
\caption{Number of $n$-body interactions in \mbox{IBM-1}.}
\label{t_numint}
\begin{center}
\begin{tabular}{lcccccc}
\hline\hline
Order&&\multicolumn{5}{c}{Number of interactions}\\
\cline{3-7}
&&total&&type BE$^a$&&type ${E_{\rm x}}^a$\\
\hline
$n=0$ &&  \phantom{0}1 && 1 &&  \phantom{0}0\\
$n=1$ &&  \phantom{0}2 && 1 &&  \phantom{0}1\\
$n=2$ &&  \phantom{0}7 && 2 &&  \phantom{0}5\\
$n=3$ && 17 && 7 && 10\\
\hline\hline
\multicolumn{7}{l}{$^a$See text for explanation.}
\end{tabular}
\end{center}
\end{table}
This analysis can be extended to higher-order interactions.
The number of possible interactions at each order $n$
is summarized in Table~\ref{t_numint} up to $n=3$.
Some of these interactions contribute to the binding energy
but do not influence the excitation spectrum of a nucleus,
as indicated with `BE' in the table.
The remaining interactions, listed under `${E_{\rm x}}$',
affect also the relative energies of the eigenstates.

\subsection{Geometric interpretation}
\label{ss_geometry}
Before entering the discussion of symmetries,
a brief discussion of the geometric interpretation is in order,
which can be obtained
by means of coherent (or intrinsic) states~\cite{Ginocchio80,Dieperink80,Bohr80}.
For the \mbox{IBM-1} the coherent state is of the form
\begin{equation}
|N_{\rm b};\alpha_\mu\rangle\propto
\Gamma(\alpha_\mu)^{N_{\rm b}}|{\rm o}\rangle,
\label{e_coherent1}
\end{equation}
where $\alpha_\mu$ are five complex variables in the expression
\begin{equation}
\Gamma(\alpha_\mu)=s^\dag+\sum_{\mu=-2}^{+2}\alpha_\mu d^\dag_\mu.
\label{e_coherent2}
\end{equation}
The $\alpha_\mu$ have the interpretation of quadrupole shape variables
and their associated conjugate momenta,
analogous to those introduced in the droplet model of the nucleus~\cite{Rainwater50,Bohr52,Bohr53}.
The real part of the $\alpha_\mu$ can be related to three Euler angles $\{\theta,\psi,\phi\}$,
which define the orientation of an intrinsic frame of reference,
and two variables, $\beta$ and $\gamma$,
that parametrize the intrinsic shape of the nuclear surface.
In terms of the latter variables the state~(\ref{e_coherent2}) is rewritten as
\begin{equation}
\Gamma(\beta,\gamma)=
s^\dag+
\beta\left[\cos\gamma d^\dag_0
+\sqrt{\frac 1 2}\sin\gamma(d^\dag_{-2}+d^\dag_{+2})\right].
\label{e_coherent3}
\end{equation}
The calculation of the expectation value of an operator in the coherent state~(\ref{e_coherent1})
leads to a function of $N_{\rm b}$, $\beta$, and $\gamma$.
The \mbox{IBM-1} Hamiltonian~(\ref{e_ham}) can be converted in this way
into a total-energy surface $E(\beta,\gamma;N_{\rm b},\epsilon,\tilde v,\dots)$,
where $\epsilon,\tilde v,\dots$ is a short-hand notation for the complete set
of parameters in the Hamiltonian.

The study of the energy surface $E(\beta,\gamma;N_{\rm b},\epsilon,\tilde v)$
has improved our understanding of the properties of the \mbox{IBM-1}
in two important ways.
First, it was instrumental in showing
that the three symmetry limits of the model, to be discussed below,
have counterparts that are also known from the geometric model of the nucleus~\cite{BM75}.
Establishing the correspondence between the IBM and the geometric model
was, in fact, one of the major achievements
in the early days of the model~\cite{Ginocchio80,Dieperink80,Bohr80}.
Secondly, the energy surface
was studied from the point of view of catastrophe theory~\cite{Gilmore81},
with the single-boson energies $\epsilon$ and boson-boson interactions $\tilde v$
viewed as control parameters
that determine the minima, saddle points etc.\ of $E(\beta,\gamma;N_{\rm b},\epsilon,\tilde v)$.
This problem was worked out for the most general \mbox{IBM-1} Hamiltonian
with up to two-body interactions~\cite{LopezMoreno96}
and also in the context of the classical Landau theory of phase transitions~\cite{Jolie02,Casten06}.
It has given rise in recent years to a flurry of activity,
which can be characterized as the study of quantum phase transitions in nuclei
(see, e.g., the review~\cite{Cejnar10}).

\subsection{Dynamical symmetries}
\label{ss_ds}
The numerical solution of the eigenvalue problem
associated with the \mbox{IBM-1} Hamiltonian~(\ref{e_ham})
can be obtained in all cases of interest,
that is, for values of $N_{\rm b}$ 
corresponding to numbers of valence pairs occurring in nuclei
and for up to three-body interactions between the bosons.
In addition, the interacting-boson problem can be solved analytically
for certain boson energies and boson-boson interactions,
and these solutions and their associated group-theoretical properties
are by now well understood~\cite{Castanos79}.
Three different analytical solutions (also known as `limits') exist:
the vibrational U(5)~\cite{Arima76},
the rotational SU(3)~\cite{Arima78},
and the $\gamma$-unstable SO(6) limit~\cite{Arima79}.
They can be summarized in a lattice of algebras of the form
\begin{equation}
{\rm U}(6)\supset
\left\{\begin{array}{c}
{\rm U}(5)\supset{\rm SO}(5)\\
{\rm SU}_\pm(3)\\
{\rm SO}_\pm(6)\supset{\rm SO}(5)
\end{array}\right\}
\supset{\rm SO}(3).
\label{e_lattice}
\end{equation}
The algebras SU(3) and SO(6) have two different realizations
depending on phase choices for the $s$ and $d$ bosons~\cite{Shirokov98},
as indicated by the $\pm$ subscripts.
In the following both algebras ${\rm SU}_\pm(3)$ are considered---they
correspond prolate and oblate shapes---whereas
only ${\rm SO}_+(6)$ is needed, henceforth denoted as SO(6).

The interpretation of the lattice~(\ref{e_lattice}) is as follows.
If the Hamiltonian can be written in terms of Casimir operators
associated with a chain of nested algebras,
then the eigenvalue problem can be solved analytically,
the quantum numbers associated with the different algebras are conserved,
and eigenfunctions are independent of the parameters in the Hamiltonian.
The underlying reason is that the Hamiltonian in that case
can be written as a sum of commuting operators
and that, as a consequence, the associated quantum numbers are conserved.
The three limits can therefore be summarized with a chain of nested algebras
and their associated quantum numbers.
For the \mbox{IBM-1} they are
\begin{align}
&\begin{array}{ccccccccc}
{\rm U}(6)&\supset&{\rm U}(5)&\supset&{\rm SO}(5)&
\supset&{\rm SO}(3)&\supset&{\rm SO}(2)\\
\downarrow&&\downarrow&&\downarrow&&\downarrow&&\downarrow\\[0mm]
[N_{\rm b}]&&n_d&&\upsilon&&\nu_\Delta L&&M
\end{array},
\nonumber\\[1ex]
&\begin{array}{ccccccc}
{\rm U}(6)&\supset&{\rm SU}_\pm(3)&\supset&{\rm SO}(3)&
\supset&{\rm SO}(2)\\
\downarrow&&\downarrow&&\downarrow&&\downarrow\\[0mm]
[N_{\rm b}]&&(\lambda_\pm,\mu_\pm)&&KL&&M
\end{array},
\nonumber\\[1ex]
&\begin{array}{ccccccccc}
{\rm U}(6)&\supset&{\rm SO}(6)&\supset&{\rm SO}(5)&
\supset&{\rm SO}(3)&\supset&{\rm SO}(2)\\
\downarrow&&\downarrow&&\downarrow&&\downarrow&&\downarrow\\[0mm]
[N_{\rm b}]&&\sigma&&\upsilon&&\nu_\Delta L&&M
\end{array}.
\label{e_clas}
\end{align}
The $N_{\rm b}$ bosons,
which can be in an $s$ or a $d$ state,
must transform symmetrically under U(6),
as indicated with the square brackets $[N_{\rm b}]$.
The allowed values for the labels of the subalgebras appearing in the lattice~(\ref{e_lattice})
then follow from standard group-theoretical reduction rules~\cite{Iachello06}.
The quantum numbers for ${\rm SU}_-(3)$ and ${\rm SU}_+(3)$
are not identical but are obtained from each other under the interchange $\lambda\leftrightarrow\mu$,
equivalent to a particle-hole transformation.
In the following only the ${\rm SU}_-(3)$ labels are needed,
henceforth denoted for simplicity as $(\lambda,\mu)$.
Note also the presence of the additional labels $K$ and $\nu_\Delta$,
which are needed to distinguish repeated angular momenta $L$
in a single SO(5) or SU(3) representation.

The preceding discussion defines the concept of a dynamical symmetry,
which has received particular attention in the context of the IBM~\cite{Iachello87}.
However, even a simplified \mbox{IBM-1} Hamiltonian reserves many further surprises
when it comes to symmetries, as will be shown in Sects.~\ref{s_pds} and~\ref{s_qds}.

\subsection{Graphical illustration}
\label{ss_grap}
\begin{figure*}
\centering
\includegraphics[width=5.7cm]{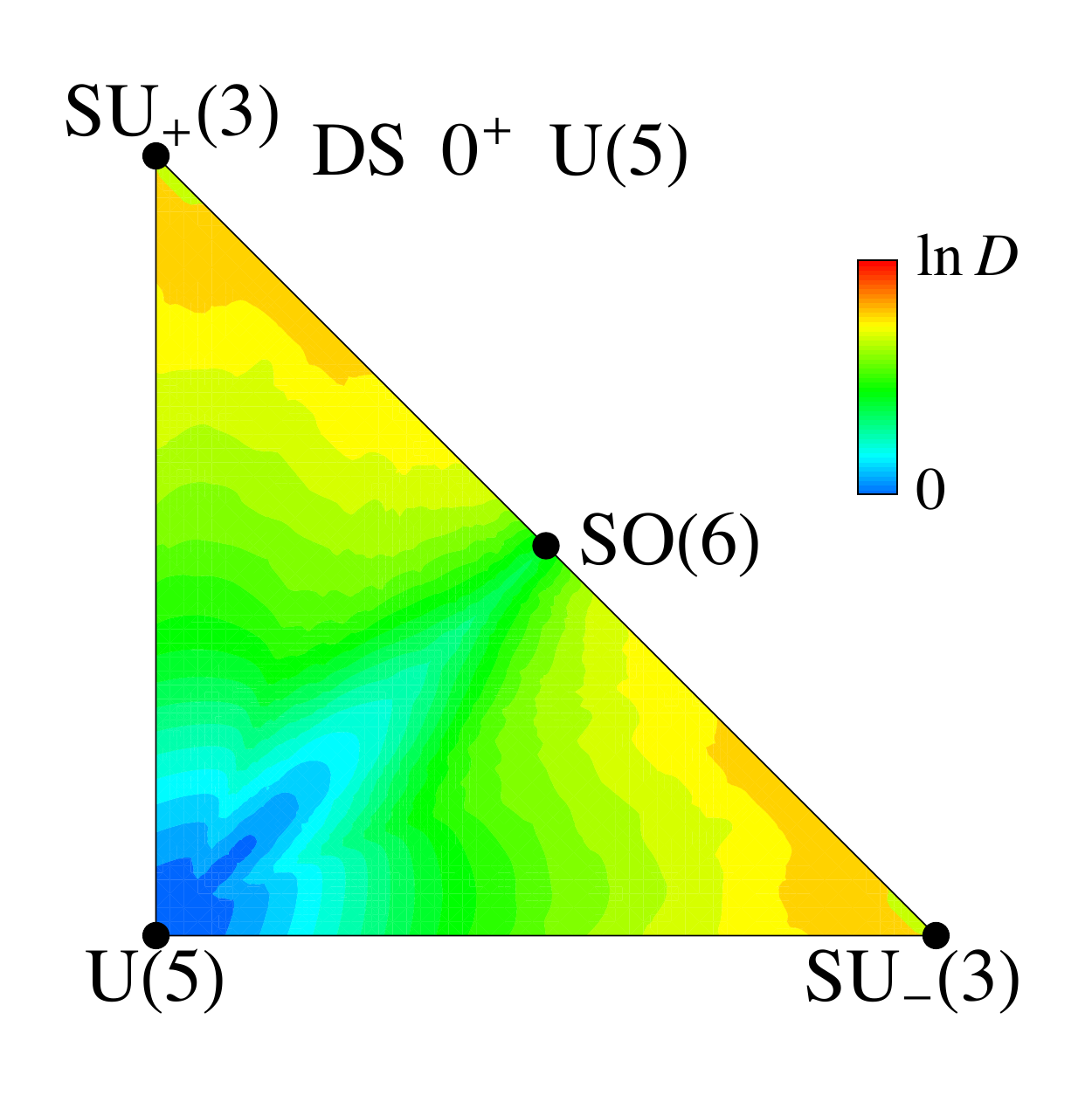}
\includegraphics[width=5.7cm]{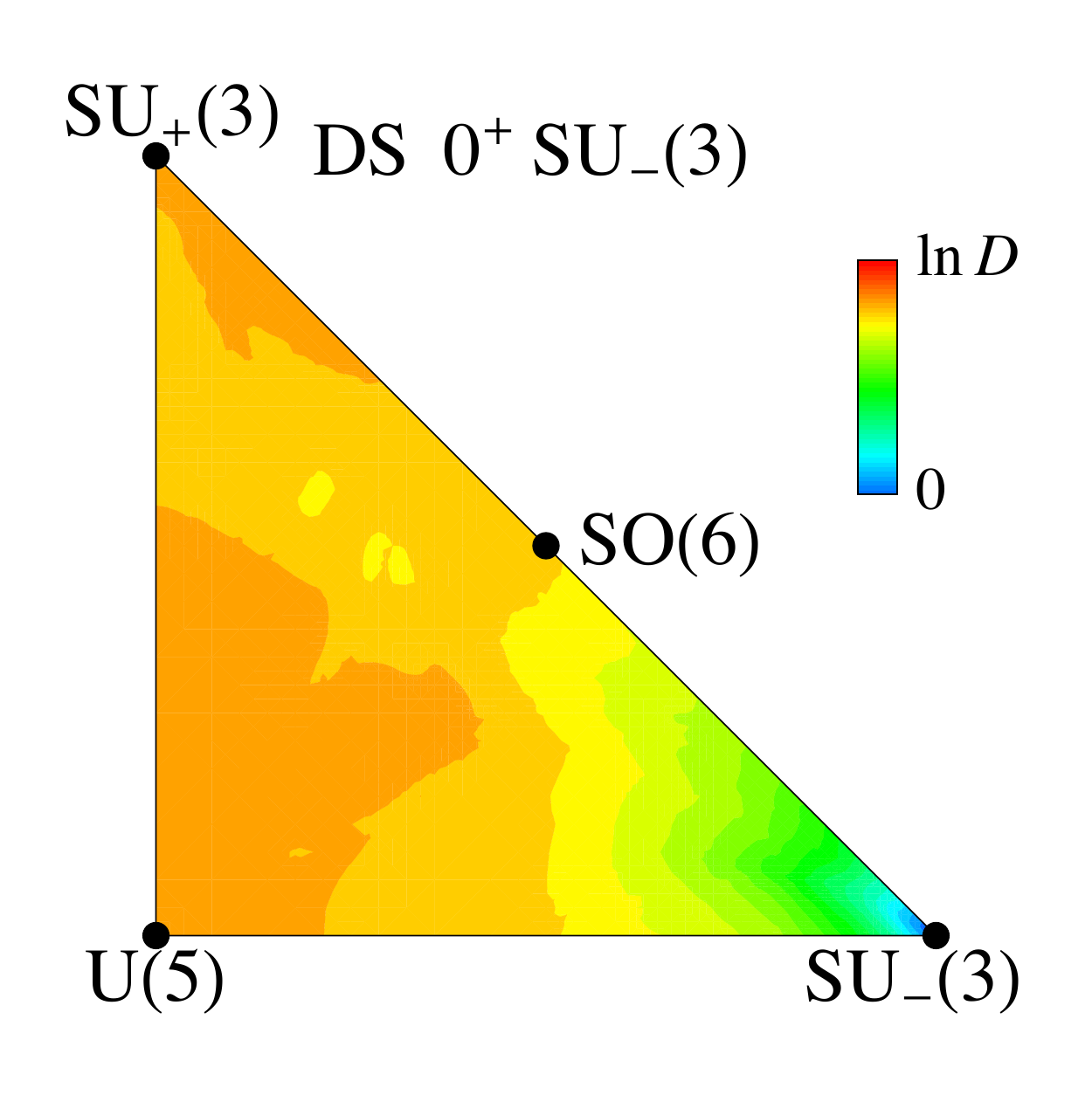}
\includegraphics[width=5.7cm]{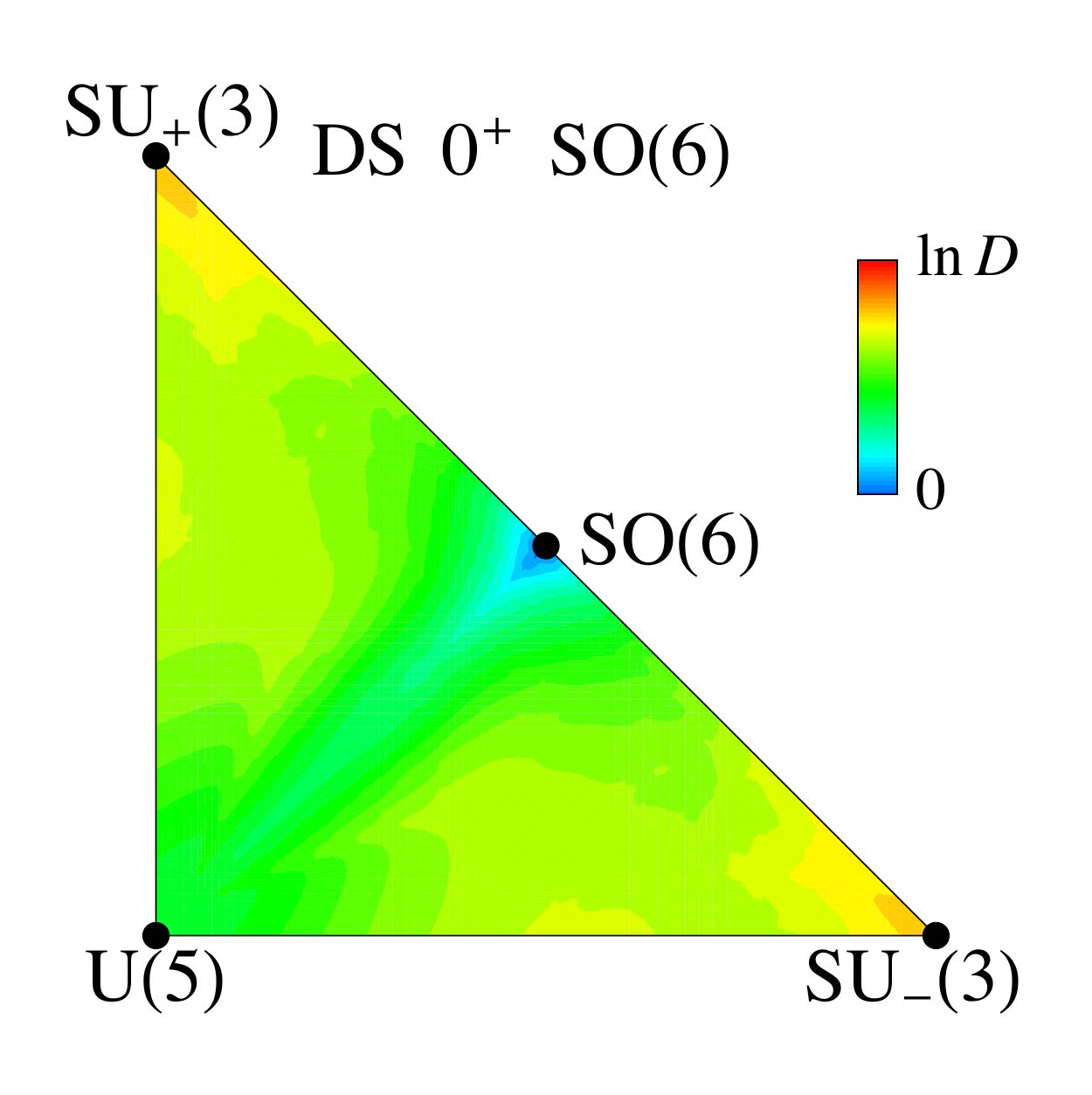}
\caption{Illustration of the three dynamical symmetries of the \mbox{IBM-1}.
The plots show the wave-function entropy for the ECQF Hamiltonian~(\ref{e_ecqf}),
in the three different bases U(5), ${\rm SU}_-(3)$, and SO(6) (left, middle, and right),
averaged over all eigenstates with angular momentum $L=0$
and boson number $N_{\rm b}=15$.}
\label{f_wfeds}
\end{figure*}
The property of dynamical symmetry can be displayed in a graphical fashion.
To explain the procedure,
consider the \mbox{IBM-1} Hamiltonian,
not in its full complexity of Eq.~(\ref{e_ham}),
but a simplified version of it,
known as the Hamiltonian of the extended consistent-$Q$ formalism (ECQF)~\cite{Warner83,Lipas85},
which reads
\begin{equation}
\hat H_{\rm ECQF}=
\omega\left[
(1-\xi)\hat n_d-\frac{\xi}{4N_{\rm b}}\hat Q^\chi\cdot\hat Q^\chi
\right],
\label{e_ecqf}
\end{equation}
where $\hat n_d$ is an operator that counts the number of $d$ bosons
and $\hat Q^\chi_\mu$ is the quadrupole operator of the model containing a parameter $\chi$,
\begin{equation}
Q^\chi_\mu=
[s^\dag\tilde d+d^\dag s]^{(2)}_\mu+\chi[d^\dag\tilde d]^{(2)}_\mu.
\label{e_quad}
\end{equation}
The eigenfunctions of the ECQF Hamiltonian
do not depend on the overall scale $\omega$ but only on $\xi$ and $\chi$,
which are therefore the structural parameters of the problem.
The parameter $\xi$ ranges from 0,
where $\hat H_{\rm ECQF}$ reduces to $\hat n_d$,
the linear Casimir operator of U(5),
to 1, where it reduces to the quadrupole term $\hat Q^\chi\cdot\hat Q^\chi$.
The latter is a combination
of quadratic Casimir operators of ${\rm SU}_\pm(3)$ and SO(3) for $\chi=\pm\sqrt{7}/2$
while for $\chi=0$ it is (up to a constant) the quadratic Casimir operator of ${\rm SO}(6)$.
A convenient range of the parameters is therefore
$0\leq\xi\leq1$ and $-\sqrt{7}/2\leq\chi\leq+\sqrt{7}/2$,
which allows to attain the U(5), ${\rm SU}_\pm(3)$, and SO(6) dynamical symmetries.
The parameter space of the ECQF Hamiltonian
can be represented on a so-called Casten triangle~\cite{Casten81},
with each point corresponding to a given $(\xi,\chi)$.

The symmetry properties of a given Hamiltonian
can be probed with use of a property called `wave-function entropy'~\cite{Cejnar98}.
For any eigenstate $|k\rangle$ of the Hamiltonian
that can be expanded in a basis $\{|i\rangle,i=1,\dots,D\}$ with components $\alpha^k_i$,
\begin{equation}
|k\rangle=\sum_{i=1}^D\alpha^k_i|i\rangle,
\label{e_expan1}
\end{equation}
the wave-function entropy is defined as
\begin{equation}
-\sum_{i=1}^D(\alpha^k_i)^2\ln(\alpha^k_i)^2.
\label{e_wfe1}
\end{equation}
The wave-function entropy of a set $\cal S$ of eigenstates of the Hamiltonian
is defined as the sum
\begin{equation}
-\frac{1}{|{\cal S}|}\sum_{k\in{\cal S}}\left(\sum_{i=1}^D(\alpha^k_i)^2\ln(\alpha^k_i)^2\right),
\label{e_wfe2}
\end{equation}
where $|{\cal S}|$ is the cardinality of the set $\cal S$,
that is, the number of eigenstates considered in the set,
such that the quantity~(\ref{e_wfe2}) represents the average wave-function entropy per eigenstate. 
It is clear from the definition that wave-function entropy depends on the basis $|i\rangle$,
which in the \mbox{IBM-1} can be taken as U(5), ${\rm SU}_\pm(3)$, or SO(6).
The property of interest here is that a vanishing wave-function entropy~(\ref{e_wfe2})
implies a dynamical symmetry. 
For example, all eigenstates $|k\rangle$ of an SO(6) Hamiltonian
have vanishing wave-function entropy in the SO(6) basis:
for each eigenstate one component $\alpha^k_i$ equals 1 and all others are 0.
However, the same SO(6) Hamiltonian
has a non-zero wave-function entropy in the U(5) or ${\rm SU}_\pm(3)$ basis,
where the SO(6) eigenstates have a fragmented structure.
The extent of this fragmentation is measured by the wave-function entropy---the
higher it is, the more fragmentation occurs.
The maximal value of the wave-function entropy
is obtained if, in a given basis of dimension $D$,
the eigenstate is completely fragmented with equal components $\pm D^{-1/2}$.
The wave-function entropy in that case reaches the value of $\ln D$.

Figure~\ref{f_wfeds} shows the wave-function entropy,
on a scale from 0 to its maximum value $\ln D$,
in the three different bases U(5), ${\rm SU}_-(3)$, and SO(6)
for all eigenstates of the ECQF Hamiltonian~(\ref{e_ecqf})
with angular momentum $L=0$ and boson number $N_{\rm b}=15$.
As argued above, wave-function entropy can be considered as a measure of dynamical symmetry
and vanishes when all quantum numbers of the basis are conserved for all eigenstates.
Therefore, a blue region (low wave-function entropy) is found
around the vertex that corresponds to the basis used to compute the wave-function entropy.
It is seen that the wave-function entropy in the bases U(5) and SO(6)
is reflection symmetric with respect to the axis U(5)--SO(6).
Following a similar line of argument,
it is not necessary to show the wave-function entropy in the SU$_+$(3) basis
since the resulting plot is the reflection-symmetric version
of the one obtained in the SU$_-$(3) basis.

\begin{figure}
\centering
\includegraphics[width=5.7cm]{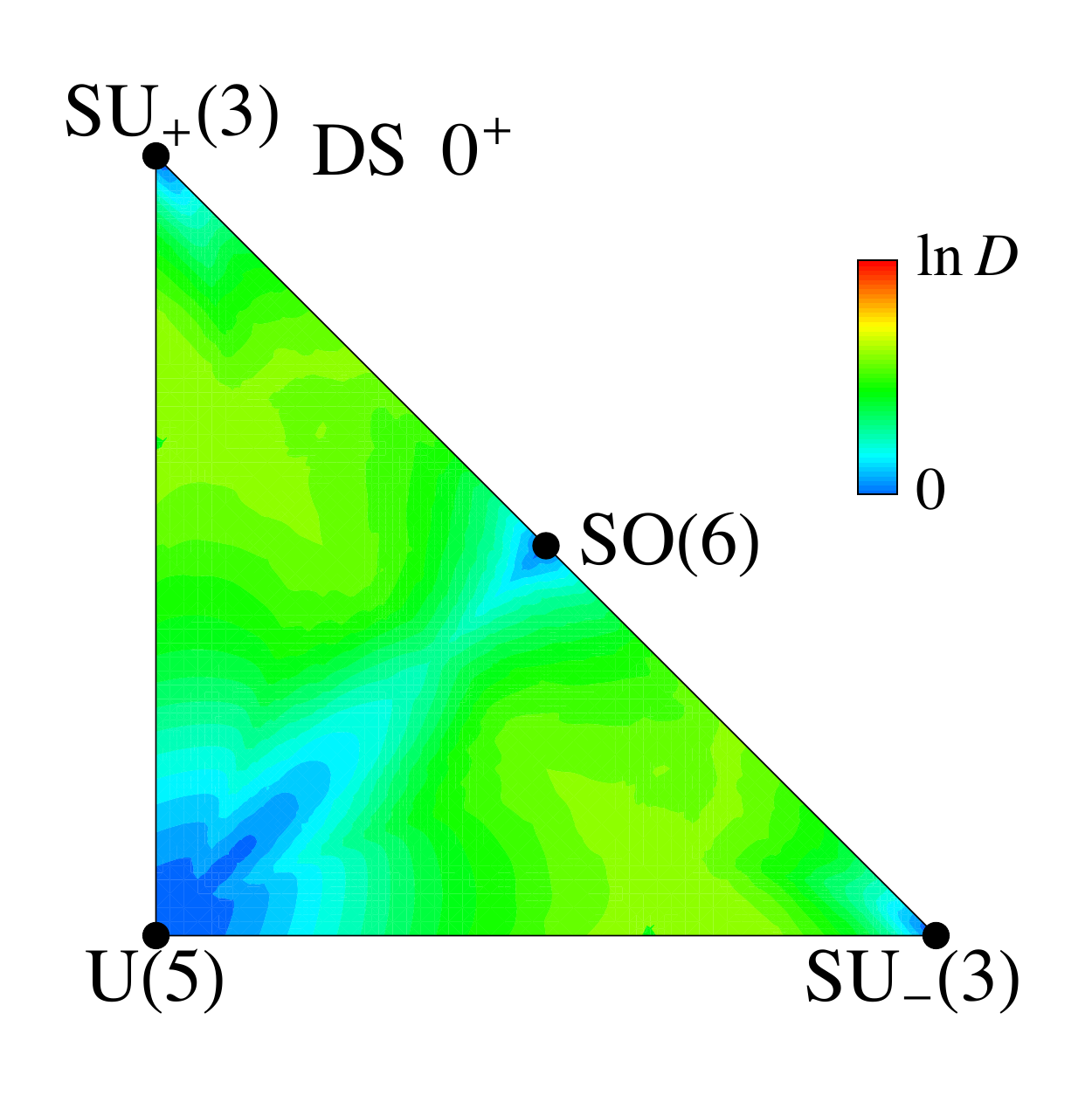}
\caption{Where in the \mbox{IBM-1} does a dynamical symmetry occur?
The plot shows the lowest value of the wave-function entropy,
as calculated in one of the four possible bases,
U(5), ${\rm SU}_-(3)$, ${\rm SU}_+(3)$, or SO(6),
for all eigenstates of the ECQF Hamiltonian~(\ref{e_ecqf})
with angular momentum $L=0$ and boson number $N_{\rm b}=15$.}
\label{f_ds}
\end{figure}
The preceding results can be conveniently summarized in a single Fig.~\ref{f_ds},
which shows the lowest value of the wave-function entropy,
as calculated in one of the four possible bases,
U(5), ${\rm SU}_-(3)$, ${\rm SU}_+(3)$, or SO(6).
This corresponds to overlaying the three plots of Fig.~\ref{f_wfeds}
and taking the minimum value at each $(\xi,\chi)$ point,
with the added requirement that
also the reflection-symmetric version of the middle plot in Fig.~\ref{f_wfeds} is considered
to account for the wave-function entropy in the ${\rm SU}_+(3)$ basis.
In the appreciation of Fig.~\ref{f_wfeds}
it should be remembered that $\ln D$ (red)
is the theoretical maximum of the wave-function entropy
and that green corresponds to about half that maximum,
that is, still considerable mixing.
Only deep-blue areas in the triangle of Fig.~\ref{f_ds}
indicate closeness to a dynamical symmetry
and, since not much blue is seen and green areas dominate,
one is tempted to conclude that most points of the triangle---and
therefore most ECQF Hamiltonians---are
not amenable to any symmetry treatment.
The main purpose of this contribution
is to show that this conclusion would be wrong.

\section{Partial dynamical symmetries}
\label{s_pds}
\begin{figure*}
\centering
\includegraphics[width=5.7cm]{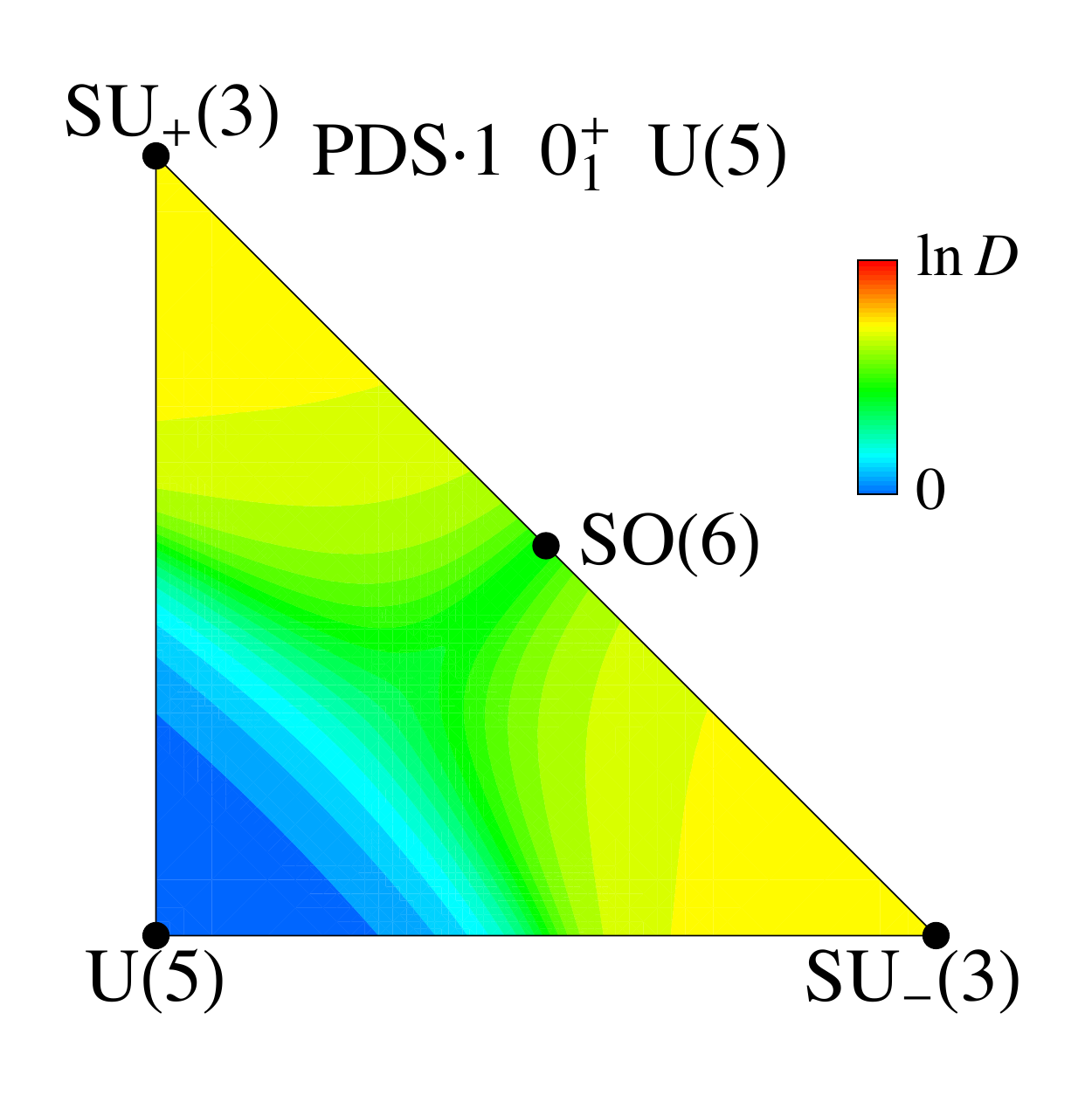}
\includegraphics[width=5.7cm]{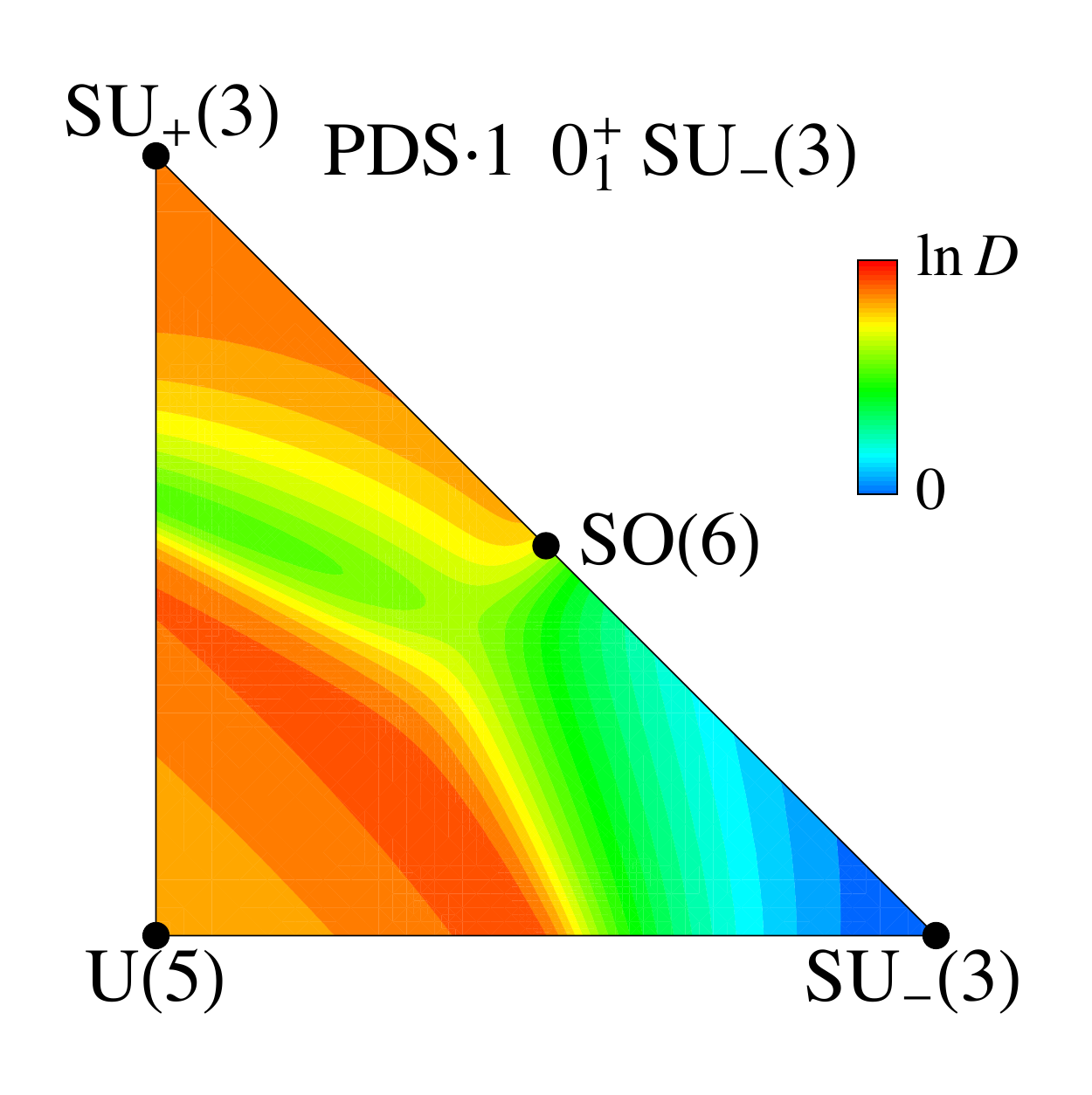}
\includegraphics[width=5.7cm]{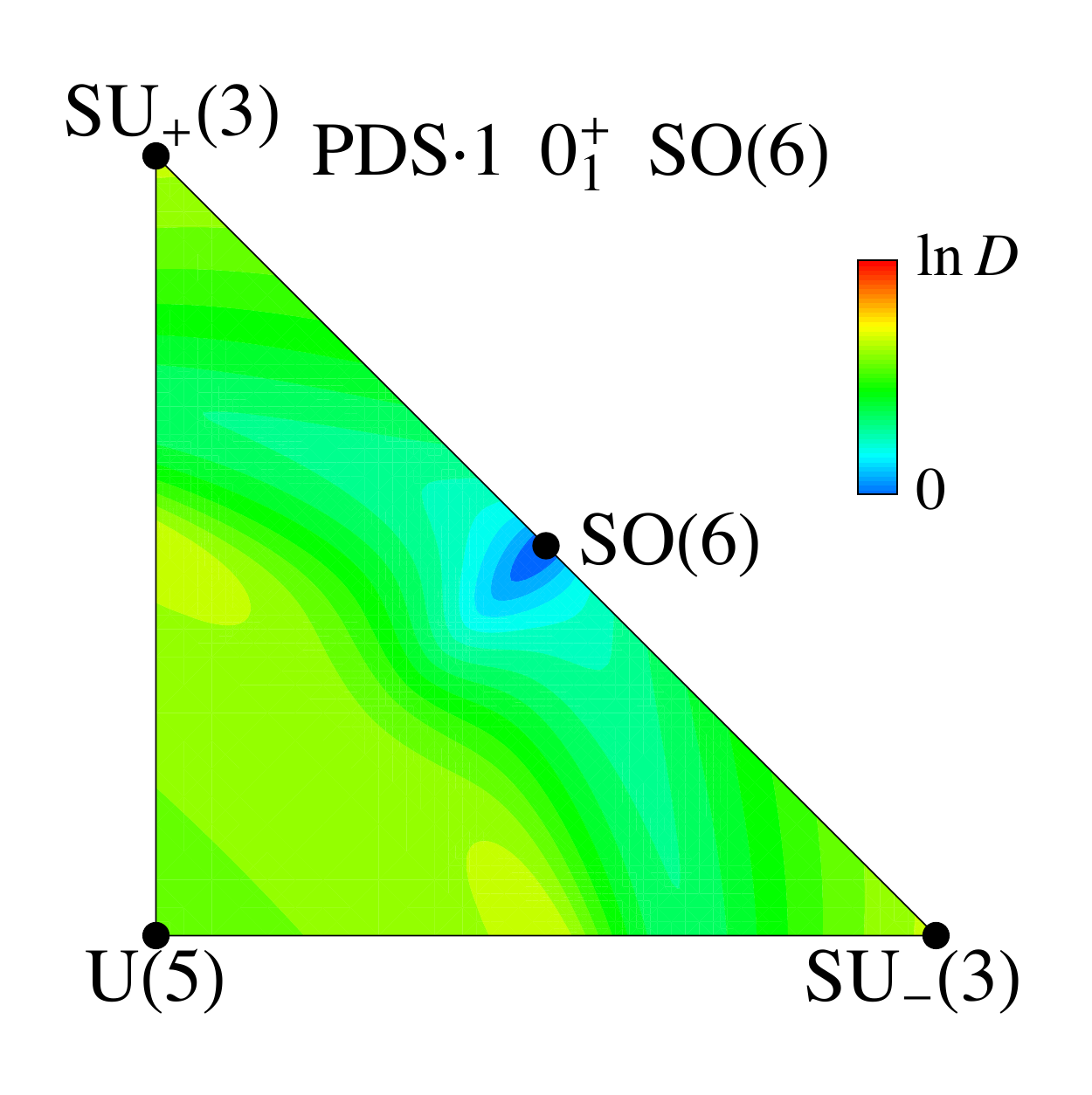}
\caption{Illustration of the three partial dynamical symmetries of the type \mbox{PDS-1} in the \mbox{IBM-1}.
The plots show the wave-function entropy of the $0^+_1$ eigenstate
with respect to all labels of the U(5), ${\rm SU}_-(3)$, and SO(6) limits
for the ECQF Hamiltonian~(\ref{e_ecqf}) with boson number $N_{\rm b}=15$.}
\label{f_wfepds1}
\end{figure*}
\begin{figure*}
\centering
\includegraphics[width=5.7cm]{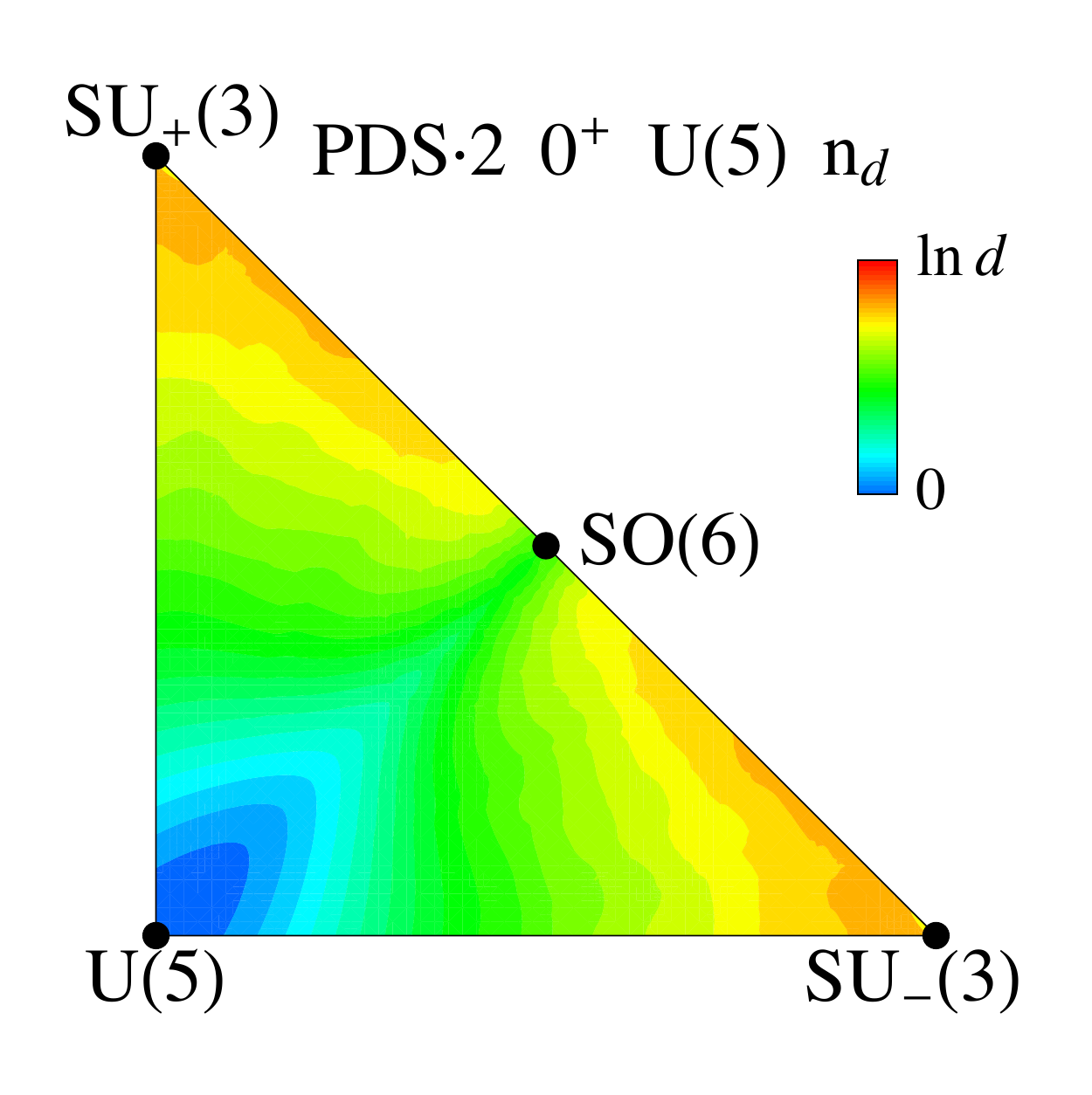}
\includegraphics[width=5.7cm]{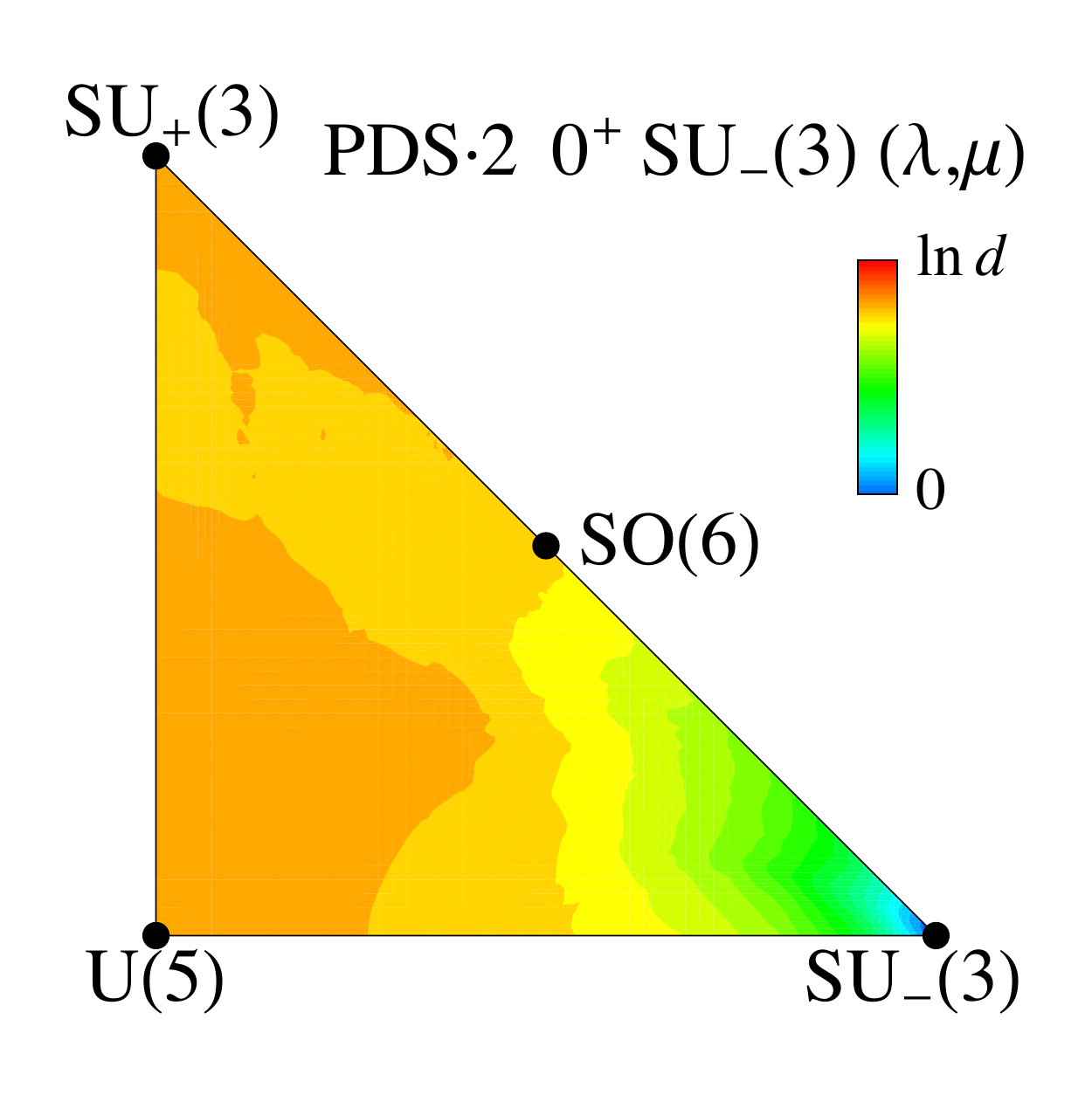}
\includegraphics[width=5.7cm]{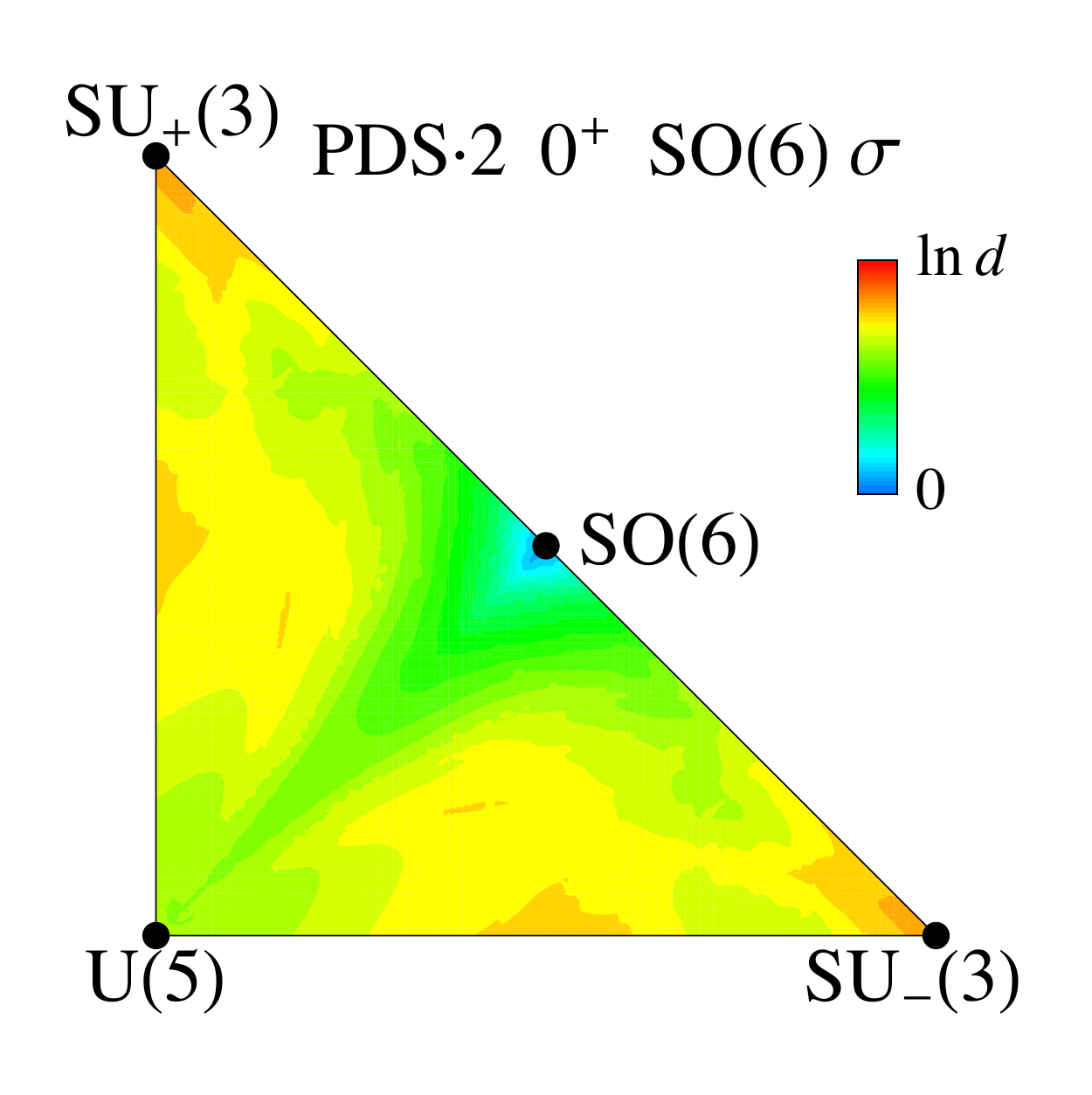}
\caption{Illustration of the three partial dynamical symmetries of the type \mbox{PDS-2} in the \mbox{IBM-1}.
The plots show the average wave-function entropy of all $0^+$ eigenstates
with respect to a single label (as indicated) of the U(5), ${\rm SU}_-(3)$, and SO(6) limits
for the ECQF Hamiltonian~(\ref{e_ecqf}) with boson number $N_{\rm b}=15$.}
\label{f_wfepds2}
\end{figure*}
\begin{figure*}
\centering
\includegraphics[width=5.7cm]{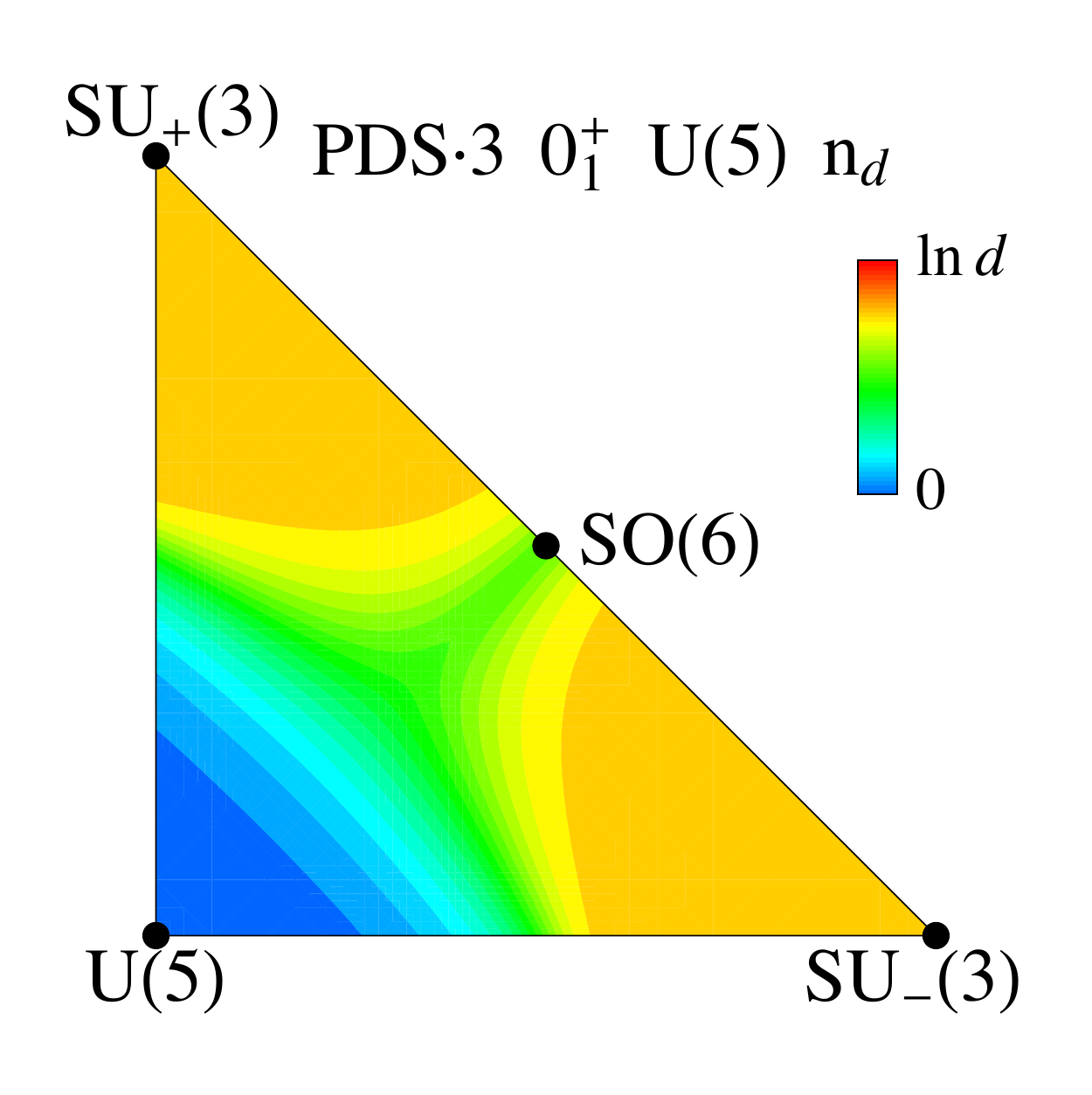}
\includegraphics[width=5.7cm]{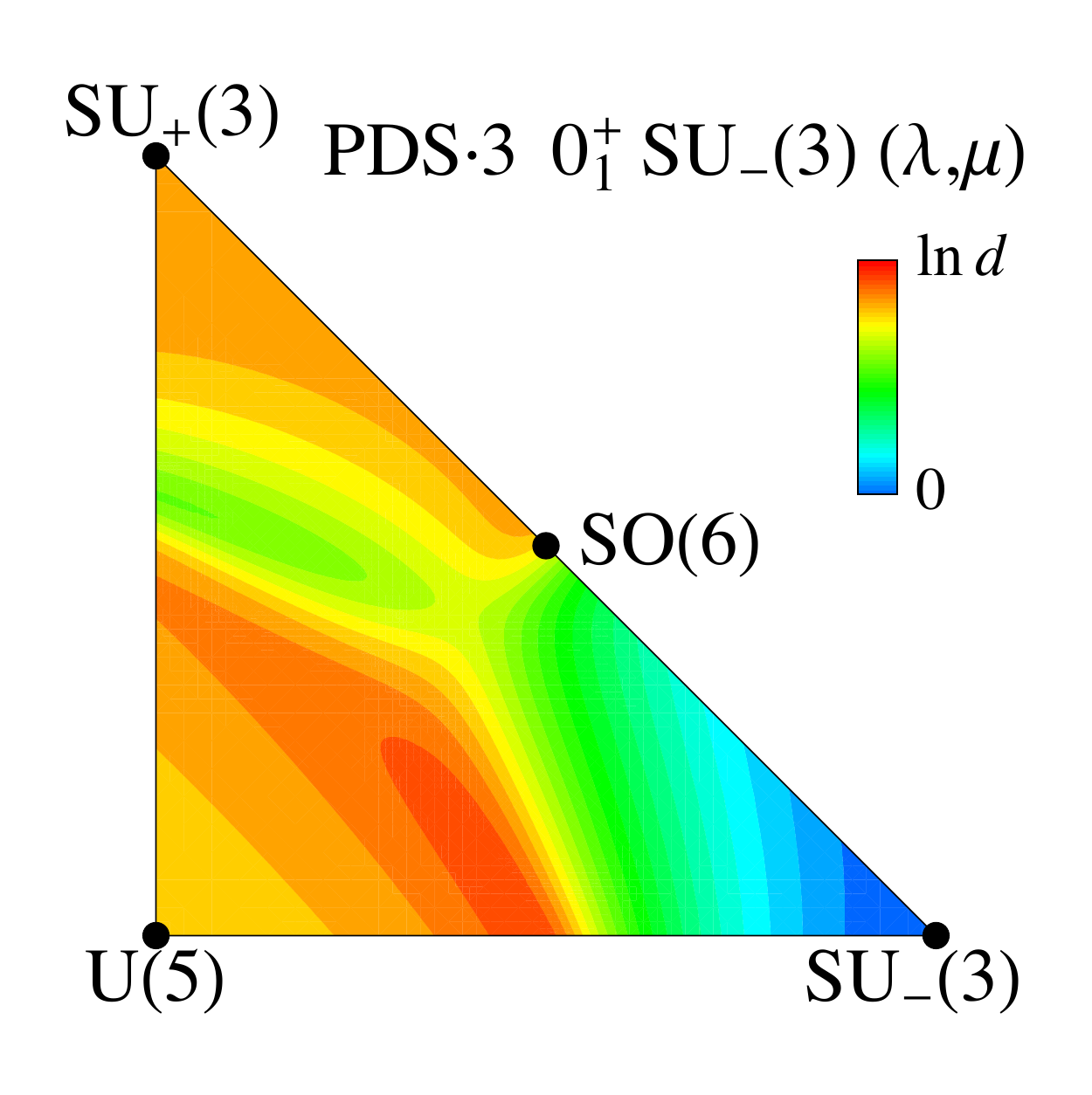}
\includegraphics[width=5.7cm]{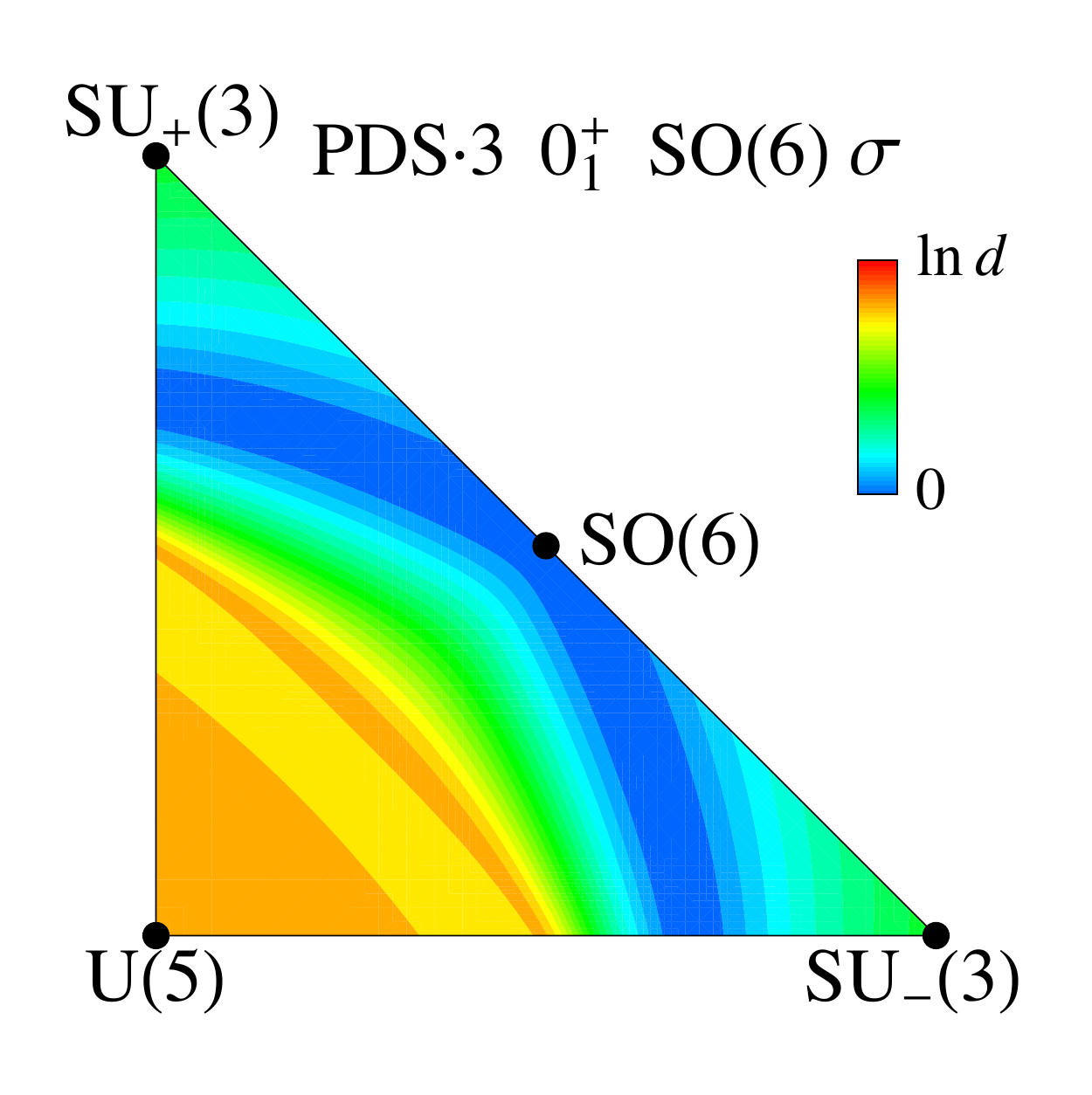}
\caption{Illustration of the three partial dynamical symmetries of the type \mbox{PDS-3} in the \mbox{IBM-1}.
The plots show the wave-function entropy of the $0^+_1$ eigenstate
with respect to a single label (as indicated) of the U(5), ${\rm SU}_-(3)$, and SO(6) limits
for the ECQF Hamiltonian~(\ref{e_ecqf}) with boson number $N_{\rm b}=15$.}
\label{f_wfepds3}
\end{figure*}
The results of Fig.~\ref{f_ds} are obtained with the expression~(\ref{e_wfe2})
where the set $\cal S$ is defined as the eigenstates
with angular momentum $L=0$ and boson number $N_{\rm b}=15$.
Similar results are obtained
if the sum is taken over {\em all} eigenstates
but for different choices of $L$ and $N_{\rm b}$.
However, one is usually interested only in eigenstates at low energy
and it makes therefore sense to restrict the set $\cal S$ to such states.
In addition, it may be that {\em some} quantum numbers
of a dynamical-symmetry classification are broken
while others are conserved.
Symmetry characteristics of this kind
can be studied by restricting the sum in Eq.~(\ref{e_wfe2})
to a subset of eigenstates of the Hamiltonian
and/or by decomposing the eigenstates onto subspaces
characterized by a single label (instead of all labels of a dynamical-symmetry chain).
Such restricted symmetries are known collectively as partial dynamical symmetries,
of which there are three different types.
In the first, \mbox{PDS-1}~\cite{Alhassid92,Leviatan96},
only eigenstates in a restricted set $\cal S$ retain all quantum numbers.
In the second type, \mbox{PDS-2}~\cite{Leviatan86,Isacker99},
all eigenstates of the \mbox{IBM-1} Hamiltonian
conserve a single label of one of the classifications~(\ref{e_clas}). 
To render the definition of the associated wave-function entropy more explicit in this case,
one decomposes each eigenstate onto subspaces
spanned by the representations of single subalgebra $G$ of U(6),
leading to the expansion
\begin{equation}
|k\rangle=\sum_{j=1}^d\sum_m\alpha^k_{jm}|jm\rangle,
\label{e_expan2}
\end{equation}
where the first sum runs over the $d$ different representations of $G$
while the second enumerates the basis states that span this representation.
With the definition of the coefficients
\begin{equation}
(\beta^k_j)^2=\sum_m(\alpha^k_{jm})^2,
\label{e_expan3}
\end{equation}
the relevant wave-function entropy of a set $\cal S$ of eigenstates can be written as
\begin{equation}
-\frac{1}{|{\cal S}|}\sum_{k\in{\cal S}}\left(\sum_{j=1}^d(\beta^k_j)^2\ln(\beta^k_j)^2\right),
\label{e_wfe3}
\end{equation}
which, by a similar argument as above, has a maximum value of $\ln d$.
Finally, the third type of partial dynamical symmetry, \mbox{PDS-3}~\cite{Leviatan02},
combines the two properties
and thus concerns a subset of eigenstates,
which is analyzed with respect to a single label.

Algorithms exist for the construction of Hamiltonians
with the required symmetry properties, \mbox{PDS-$i$},
and can be found in the review~\cite{Leviatan11}.

As before, the concept of partial dynamical symmetry
can be illustrated graphically with the wave-function entropy of the ECQF Hamiltonian~(\ref{e_ecqf}).
Figures~\ref{f_wfepds1} to~\ref{f_wfepds3} shows the results of nine different calculations,
varying the set ${\cal S}$ of eigenstates,
the choice of the label [$n_d$, $(\lambda,\mu)$, or $\sigma$],
and the basis [U(5), ${\rm SU}_-(3)$, or SO(6)],
always for angular momentum $L=0$ and boson number $N_{\rm b}=15$.
On the left-hand panel of each figure is plotted the wave-function entropy
of the $0^+_1$ eigenstate, that is, for ${\cal S}=\{|0^+_1\rangle\}$,
decomposed in the three different bases,
U(5), ${\rm SU}_-(3)$, and SO(6).
In the middle panel the wave-function entropy
is summed over all $0^+$ eigenstates
but the components $\beta^k_j$ correspond to the decomposition
onto subspaces that are characterized by a single label,
$n_d$, $(\lambda,\mu)$, or $\sigma$,
as in Eq.~(\ref{e_wfe3}).
The wave-function entropy of the $0^+_1$ ground state
with respect to a single label is shown on the right-hand panel of Figs.~\ref{f_wfepds1} to~\ref{f_wfepds3}.
The figures therefore illustrate graphically
the three types of partial dynamical symmetry \mbox{PDS-$i$}.
A remarkable result is found
as regards the conservation of the SO(6) label $\sigma$
namely, the existence of an entire band of ECQF Hamiltonians
with close to exact SO(6) symmetry in the ground state~\cite{Kremer14},
see the left-hand panel of Fig.~\ref{f_wfepds3}.

\begin{figure}
\centering
\includegraphics[width=5.7cm]{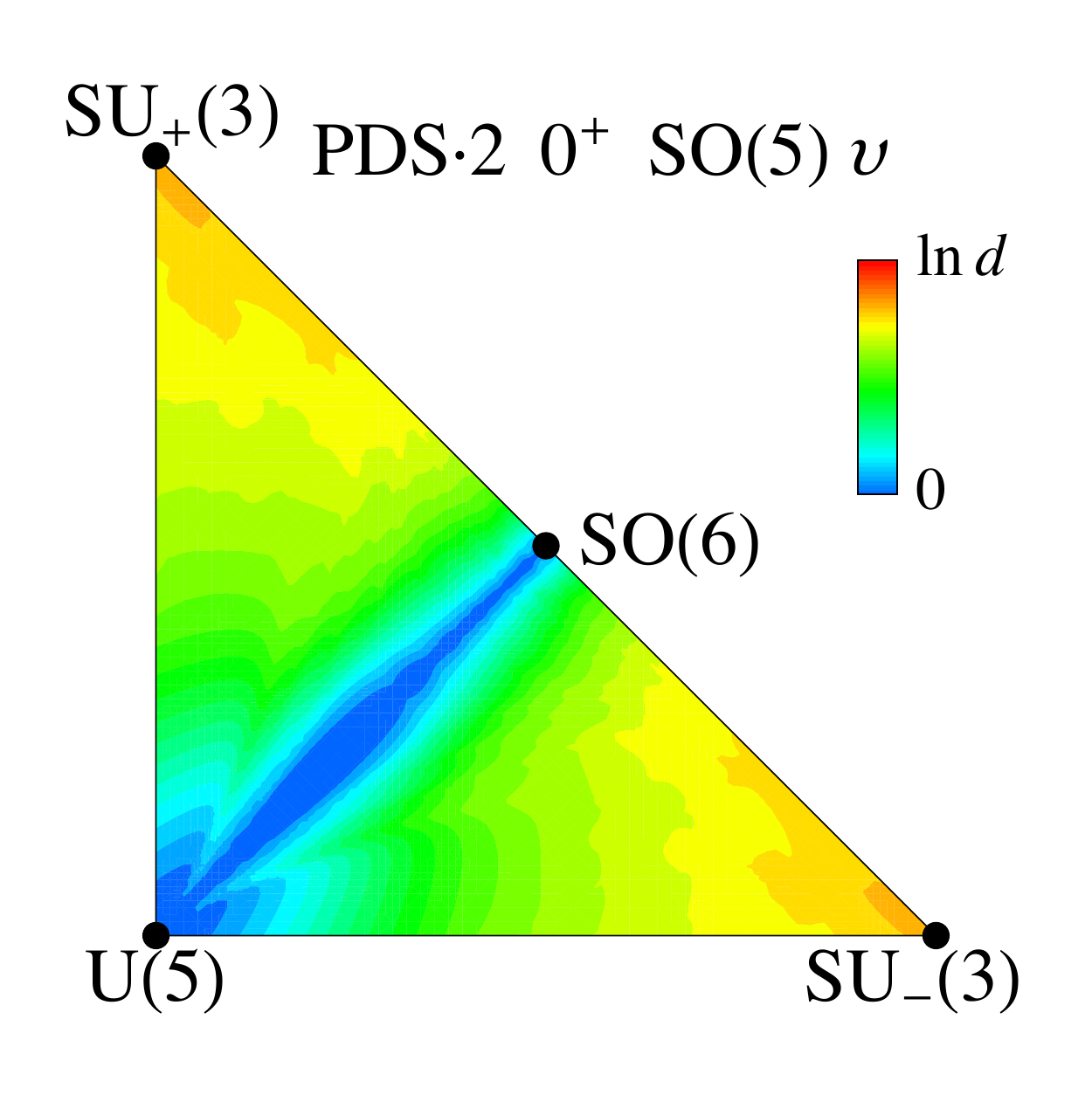}
\includegraphics[width=5.7cm]{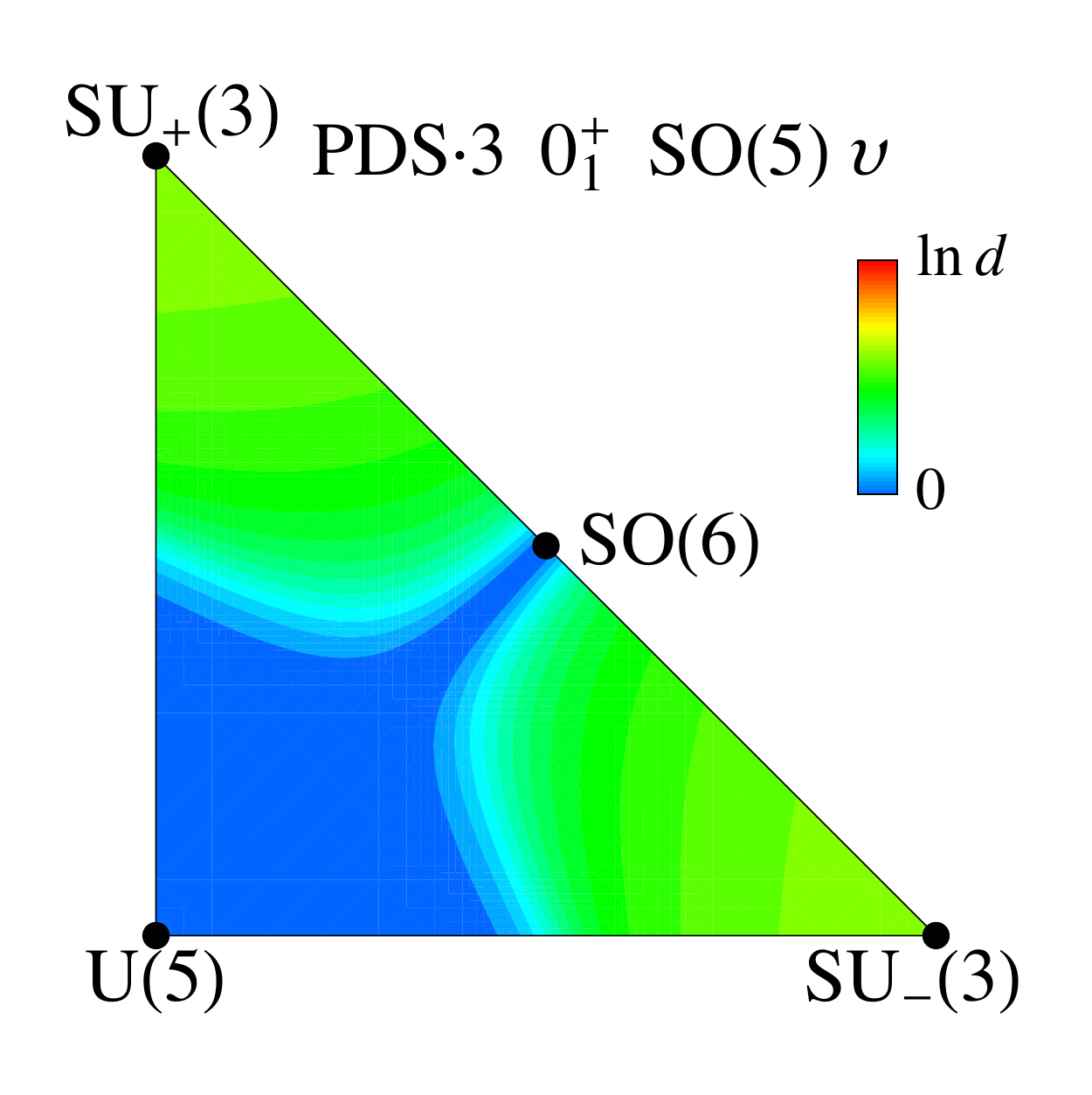}
\caption{Illustration of the SO(5) partial dynamical symmetry in the \mbox{IBM-1}.
The plots show the wave-function entropy for the eigenstates
of the ECQF Hamiltonian~(\ref{e_ecqf})
with angular momentum $L=0$ and boson number $N_{\rm b}=15$.
Top: average wave-function entropy of all $0^+$ eigenstates with respect to the SO(5) label $\upsilon$ (\mbox{PDS-2}).
Bottom: wave-function entropy of the $0^+_1$ eigenstate with respect to $\upsilon$ (\mbox{PDS-3}).}
\label{f_wfepdsso5}
\end{figure}
A partial dynamical symmetry can also be defined
with respect to the SO(5) label $\upsilon$---associated with $d$-boson seniority.
Given its single-label character,
it concerns either \mbox{PDS-2} or \mbox{PDS-3}.
The top panel in Fig.~\ref{f_wfepdsso5} shows that the conservation of the SO(5) label
is exact for the entire U(5)--SO(6) transitional Hamiltonian,
as is known since long~\cite{Leviatan86}.
Moreover and more generally, it can be shown
that the U(5)--SO(6) transitional Hamiltonian is integrable~\cite{Pan98}.
The bottom panel in Fig.~\ref{f_wfepdsso5}
illustrates the conservation of the SO(5) label in the $0^+_1$ eigenstate
and it is seen that a large area
corresponds to ECQF Hamiltonians with approximate SO(5) symmetry in the ground state.
Selection rules associated with this symmetry
can therefore be expected to have a wide validity in nuclei.

\begin{figure}
\centering
\includegraphics[width=5.7cm]{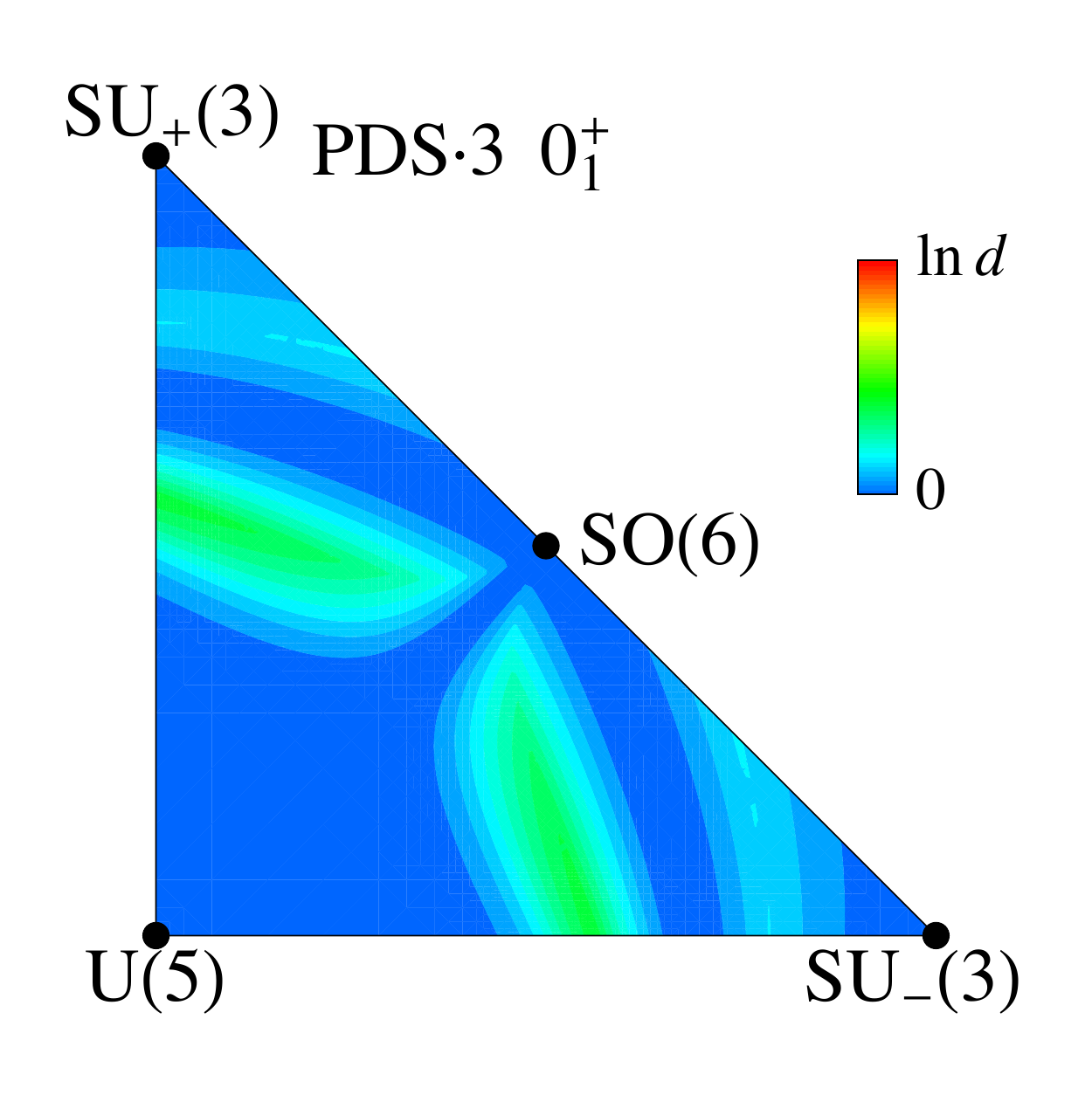}
\caption{Where in the \mbox{IBM-1} does a partial dynamical symmetry occur?
The plot shows the lowest value of the wave-function entropy,
as calculated with respect to five possible labels
$n_d$, $(\lambda,\mu)$, $(\mu,\lambda)$, $\sigma$, or $\upsilon$,
for the $0^+_1$ eigenstate of the ECQF Hamiltonian~(\ref{e_ecqf})
for boson number $N_{\rm b}=15$ (\mbox{PDS-3}).}
\label{f_pds3}
\end{figure}
These results can be summarized again in a single plot, Fig.~\ref{f_pds3},
which shows the lowest value of the wave-function entropy of the $0^+$ ground state,
calculated with respect to the U(5), ${\rm SU}_-(3)$, ${\rm SU}_+(3)$, SO(6), or SO(5) labels.
The figure illustrates that a large fraction of the parameter space of the ECQF Hamiltonian~(\ref{e_ecqf})
enjoys an approximate symmetry of one kind or another in its ground state.

\section{Quasi dynamical symmetries}
\label{s_qds}
\begin{figure*}
\centering
\includegraphics[width=5.7cm]{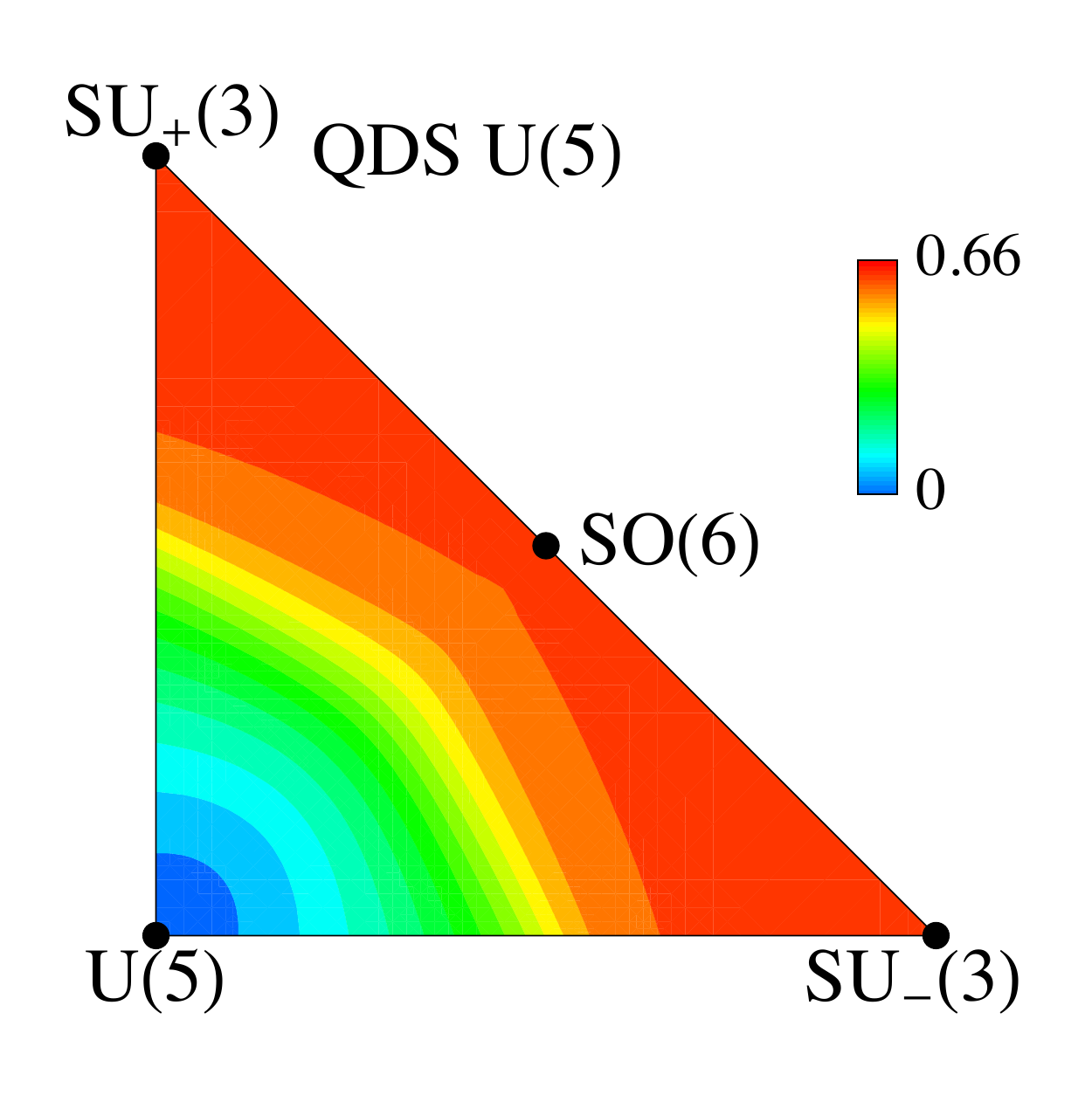}
\includegraphics[width=5.7cm]{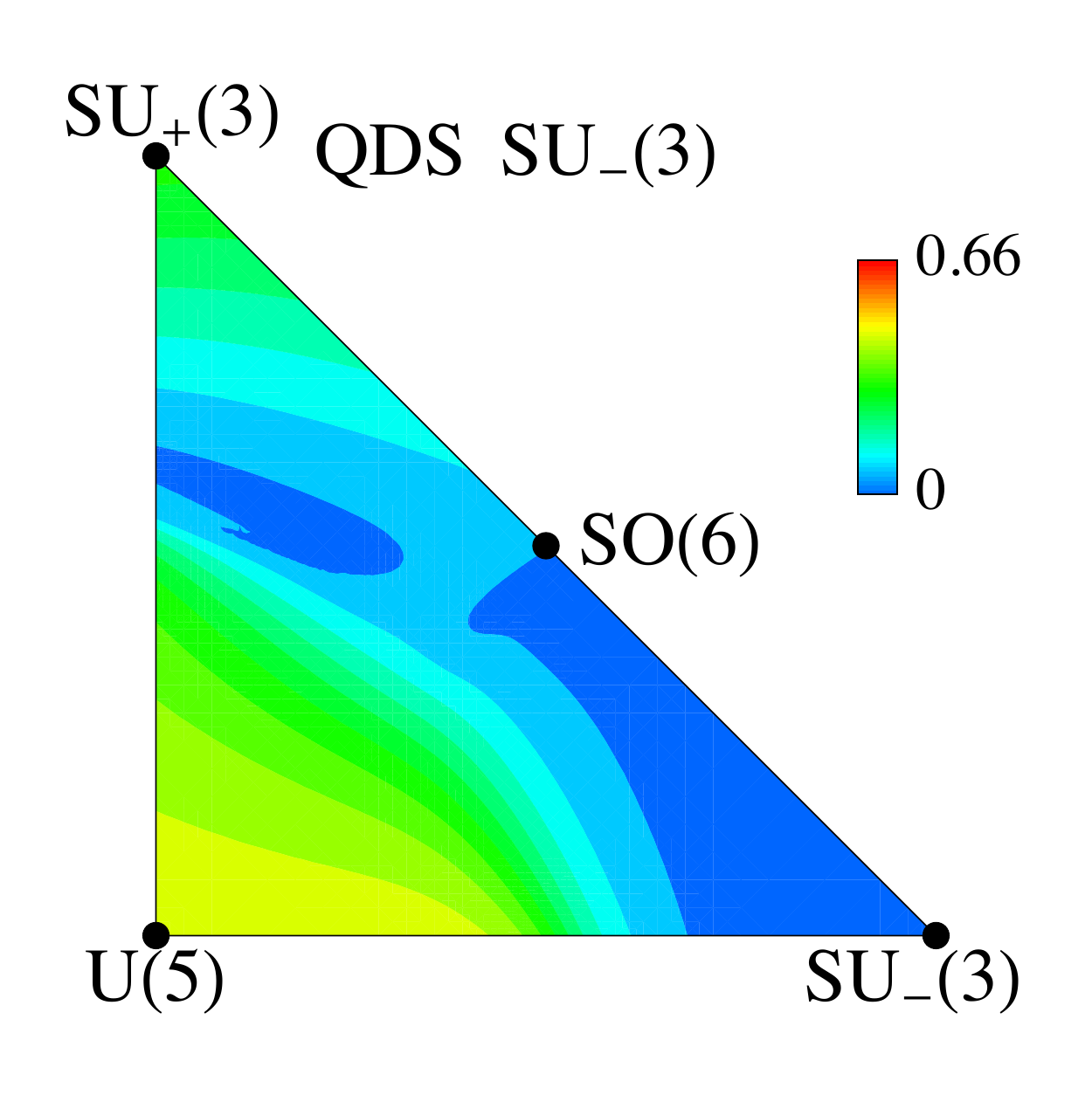}
\includegraphics[width=5.7cm]{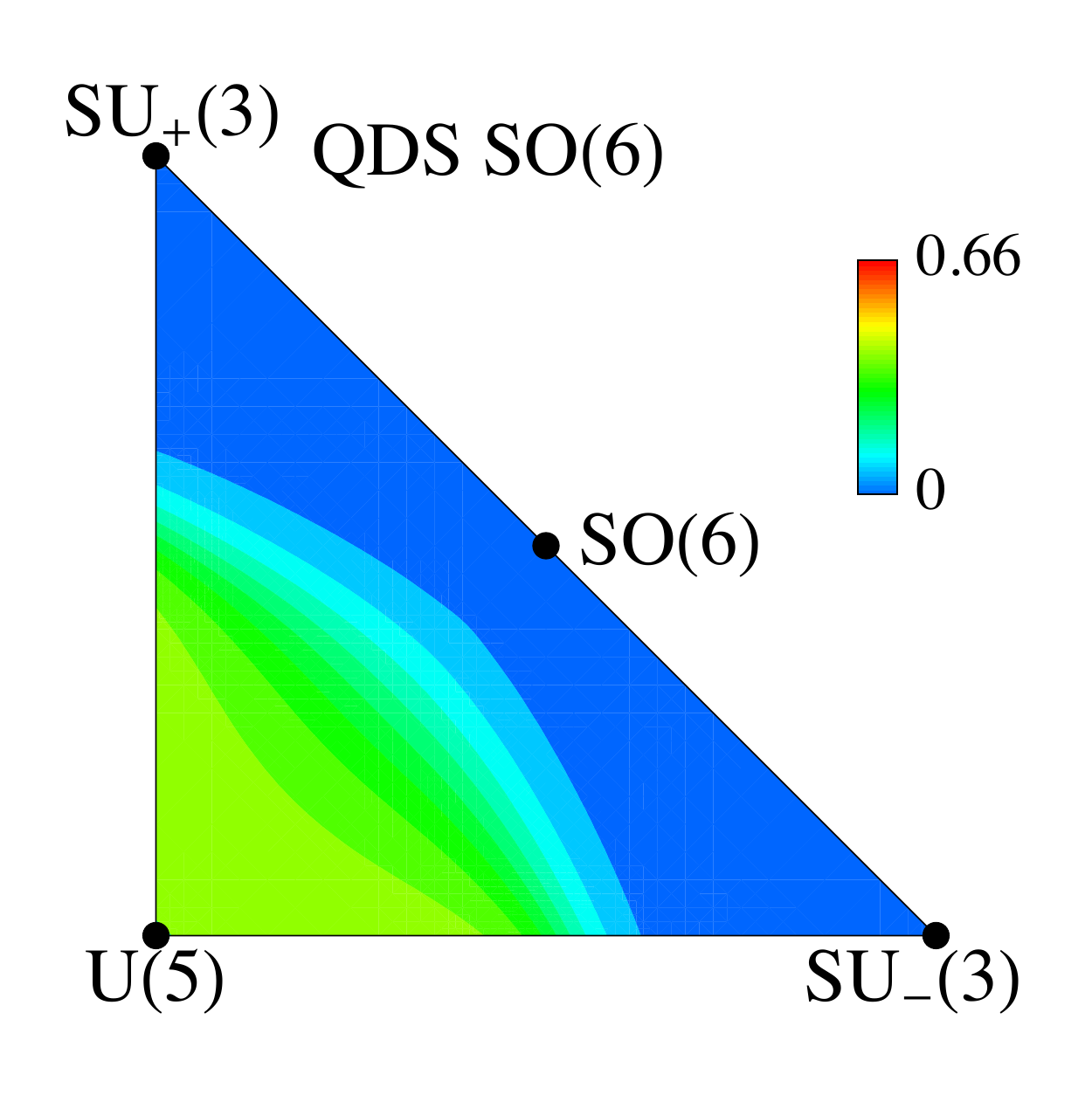}
\caption{Illustration of the three quasi dynamical symmetries of the \mbox{IBM-1}.
The plots show the quantity $\Omega_r$,
a measure of quasi dynamical symmetry as defined in the text,
for the ECQF Hamiltonian~(\ref{e_ecqf})
in the three different bases U(5), ${\rm SU}_-(3)$, and SO(6) (left, middle, and right),
for yrast eigenstates with angular momenta $L=0,2,\dots,10$
and for boson number $N_{\rm b}=15$.}
\label{f_qds1}
\end{figure*}
Quasi dynamical symmetries
constitute another extension of the concept of dynamical symmetry.
They can be given a mathematical definition in terms of embedded representations~\cite{Rowe88}.
The (admittedly loose) physical interpretation of quasi dynamical symmetries
is that observables can be consistent with a certain symmetry,
which is in fact broken in the Hamiltonian.
Typically, this situation occurs for a Hamiltonian that is transitional between two limits
and which retains, for a certain range of its parameters,
the characteristic patterns of one of those dynamical symmetries~\cite{Rowe98,Bahri00,Macek09,Bonatsos10}.
In more mathematical terms a coherent mixing of representations in a subset of eigenstates
is at the basis of this `apparent' symmetry.

The validity of a quasi dynamical symmetry
must be probed by examining the similarity in the decomposition of certain eigenstates.
A quantitative measure of quasi dynamical symmetry
can be introduced by rewriting the expansion~(\ref{e_expan1}) in yet another way,
\begin{equation}
|k\rangle\equiv|rL\rangle=\sum_{i=1}^{D_L}\alpha^{rL}_i|i\rangle,
\label{e_expan3}
\end{equation}
where $\{k\}\equiv\{rL\}$, that is, $r$ contains all labels except the angular momentum $L$.
The measure of quasi dynamical symmetry
is defined as $\Omega_r\equiv\sqrt{1-\Theta_r}$
where $\Theta_r$ is the average over pairs $L\neq L'$ of the quantities
\begin{equation}
\Theta_r^{LL'}\equiv\sum_{i=1}^{D_L}\alpha^{rL}_i\alpha^{rL'}_i.
\label{e_qds}
\end{equation}
A vanishing $\Omega_r$ therefore indicates
a perfect correlation between the expansion coefficients $\alpha^{rL}_i$
with different angular momenta $L$.
In a typical application of quasi dynamical symmetry
one wishes to probe the similarity of the structure of yrast states,
which implies the identification of $r$ with the labels of the ground-state band,
that is, for $n_d=L/2$ in U(5),
$(\lambda,\mu)=(2N,0)$ in SU(3),
and $\sigma=N$ in SO(6).
Figure~\ref{f_qds1} shows the quantity $\Omega_r$ for the ECQF Hamiltonian~(\ref{e_ecqf})
in the three different bases U(5), ${\rm SU}_-(3)$, and SO(6),
for the angular momenta $L=0,2,\dots,10$ and boson number $N_{\rm b}=15$.

It is obvious that connections exist
between the concepts of partial and dynamical symmetry.
For example, the band structure
in the wave-function entropy of the ground state with respect to $\sigma$ in Fig.~\ref{f_wfepds3}
also shows up in the SO(6) quasi dynamical symmetry of Fig.~\ref{f_qds1}.
A remarkable finding of this analysis
is that the partial conservation of one symmetry
may occur simultaneously with the coherent mixing of another, incompatible symmetry.

\begin{figure}
\centering
\includegraphics[width=5.7cm]{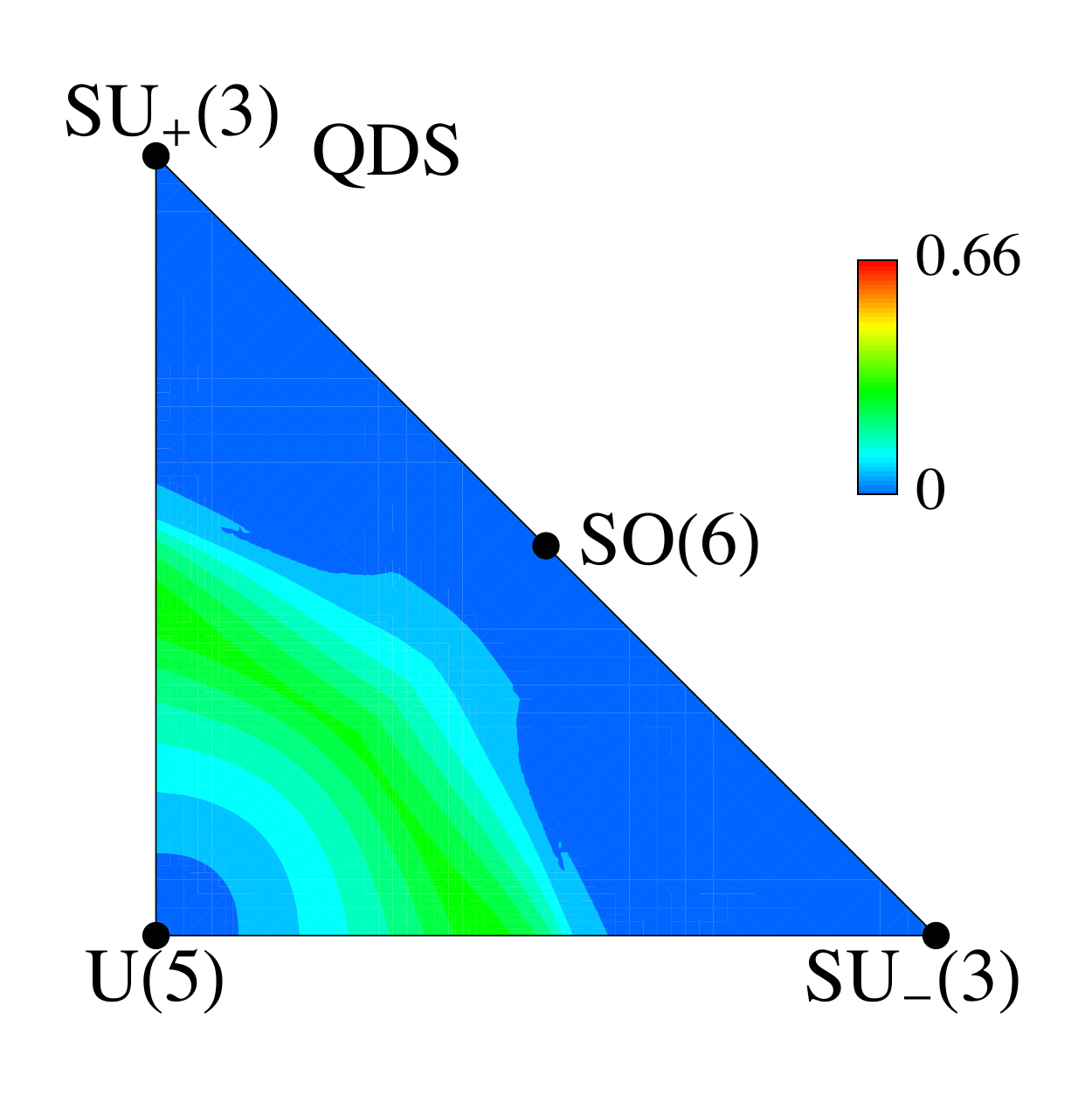}
\caption{Where in the \mbox{IBM-1} does a quasi dynamical symmetry occur?
The plot shows the lowest value of $\Omega_r$
in any of the four bases U(5), ${\rm SU}_-(3)$, ${\rm SU}_+(3)$, and SO(6),
for yrast eigenstates of the ECQF Hamiltonian~(\ref{e_ecqf})
with angular momenta $L=0,2,\dots,10$
and for boson number $N_{\rm b}=15$.
The quantity $\Omega_r$ is a measure of quasi dynamical symmetry, as defined in the text.}
\label{f_qds2}
\end{figure}
Again one can summarize these findings in a single Fig.~\ref{f_qds2},
which displays the minimum value of the measure $\Omega_r$
in any of the four bases,
U(5), ${\rm SU}_-(3)$, ${\rm SU}_+(3)$, or SO(6).
Large areas of the parameter space
are seen to be blue, that is, to display a quasi dynamical symmetry.

The contrast between the results shown
for dynamical symmetries on the one hand, Fig.~\ref{f_ds},
and those for partial and quasi dynamical symmetries on the other,
Figs.~\ref{f_pds3} and~\ref{f_qds2}, is startling.
Dynamical symmetries are restricted to small regions in the parameter space
(the blue areas in Fig.~\ref{f_ds})
and therefore are expected to have only restricted applicability in nuclei.
This is not the case for the extended concepts
of partial and quasi dynamical symmetries,
as illustrated in Figs.~\ref{f_pds3} and~\ref{f_qds2},
where large bands of blue are found in the triangle.

\section{Concluding remarks}
\label{s_conc}
Dynamical symmetries are scarce
while partial dynamical symmetries and quasi dynamical symmetries are ubiquitous.
This has been the main theme of this contribution.
It has been examined in the context of the interacting boson model
for the schematic Hamiltonian of the extended consistent-$Q$ formalism
and illustrated by a graphical representation of wave-function entropy in various bases.
In no way do these results represent the complete symmetry analysis of the IBM.
A general Hamiltonian of the interacting boson model with up to two-body interactions
allows the occurrence of exact dynamical symmetries of various partialities,
some of which are not or only approximately present
in the schematic Hamiltonian of the extended consistent-$Q$ formalism.
Also, given the composite nature of the bosons
three-body interactions between them are to be expected,
further enriching the symmetry features of the model.
It is remarkable that more than forty years after the proposal by Arima and Iachello,
the full symmetry content of the interacting boson model still remains to be uncovered.

\section{Ackowledgements}
I wish, on the occasion of the 88th anniversary of his birthday,
to express my sincere thanks to Akito Arima
for many years of stimulating discussions and his continual inspiration of my research.
Many thanks are due to Amiram Leviatan and Jos\'e-Enrique Garc\'\i a-Ramos,
in collaboration with whom many of the results reported in this contribution have been obtained.

\end{document}